\documentclass[twocolumn,showpacs,superscriptaddress,longbibliography]{revtex4-1}
\usepackage{graphicx}
\usepackage{array}
\usepackage{amsmath}
\usepackage{xcolor}
\usepackage[colorlinks=true, citecolor={blue!80!black}, urlcolor={blue!50!black}, linkcolor = {blue!80!black}]{hyperref}
\usepackage{amssymb}
\usepackage{bm}
\usepackage{color}
\usepackage{bookmark}
\usepackage{tabularx}
\usepackage{mathtools}
\usepackage{microtype}
\usepackage{hhline}

\DeclareMathOperator{\e}{e}
\DeclareMathOperator{\de}{d\!}

\DeclareMathOperator{\sgn}{sgn}

\DeclareMathOperator{\re}{Re}

\newcommand{\ket}[1]{|#1\rangle}

\newcommand{\abs}[1]{\left\vert#1\right\vert} 

\newcommand{\pd}{\partial}
\newcommand{\av}[1]{\left\langle#1\right\rangle}

\newcommand{\up}{\uparrow}
\newcommand{\down}{\downarrow}

\begin{document}

\title{Zeeman and spin-orbit effects in the Andreev spectra of nanowire junctions}
\author{B. van Heck}
\affiliation{Department of Physics, Yale University, New Haven, CT 06520, USA}
\author{J. I. V\"{a}yrynen}
\affiliation{Department of Physics, Yale University, New Haven, CT 06520, USA}
\author{L. I. Glazman}
\affiliation{Department of Physics, Yale University, New Haven, CT 06520, USA}
\date{\today}
\begin{abstract}
We study the energy spectrum and the electromagnetic response of Andreev bound states in short Josephson junctions made of semiconducting nanowires. We focus on the joint effect of Zeeman and spin-orbit coupling on the Andreev level spectra. Our model incorporates the penetration of the magnetic field in the proximitized wires, which substantially modifies the spectra. We pay special attention to the occurrence of fermion parity switches at increasing values of the field and to the magnetic field dependence of the absorption strength of microwave-induced transitions. Our calculations can be used to extract quantitative information from microwave and tunneling spectroscopy experiments, such as the recently reported measurements in Van Woerkom et al.~\cite{vanwoerkom2016}.
\end{abstract}

\maketitle

\section{Introduction}

The Josephson current flowing across a weak link between two superconductors is mediated by Andreev bound states \cite{golubov2004}, sub-gap states localized at the position of the weak link.
Recent years have witnessed the direct observation of Andreev bound states in different types of weak links \cite{pillet2010,deacon2010,zgirski2011,bretheau2013,bretheau2013b,chang2013,lee2014}, via either tunneling or microwave spectroscopy, as well as their coherent manipulation \cite{janvier2015}.
Aside from increasing our understanding of mesoscopic superconductivity, these results pave the way to the realization of novel types of qubits \cite{zazunov2003,chtchelkatchev2003,padurariu2010,padurariu2012} and superconducting circuits.
Among these results, of particular interest is the very recent microwave detection of Andreev bound states in InAs/Al nanowires \cite{vanwoerkom2016}.
Such hybrid semiconducting/superconducting systems are under intense investigation \cite{mourik2012,higginbotham2015,albrecht2015,deng2016,zhang2017} as platforms for Majorana zero modes \cite{alicea2012,beenakker2013b,leijnse2012,stanescu2013} and, eventually, topological quantum computation \cite{nayak2008,dassarma2015}.
In these devices, the study of Andreev bound states may be a prelu,de for the study of Majoranas in microwave circuits and the realization of topological qubits.

In view of these exciting applications, the measurement of the magnetic field dependence of the Andreev spectra was among the most interesting aspects of the experiment of Ref.~\cite{vanwoerkom2016}.
InAs (or InSb) nanowires are characterized by strong spin-orbit coupling and large g-factors: both are necessary ingredients to reach the topological phase with Majorana bound states which is predicted to occur at high magnetic field and low electron density \cite{lutchyn2010,oreg2010}.
Spectroscopic studies of nanowire Josephson junctions, even if performed in the topologically trivial phase, can bring quantitative understanding about the interplay of Zeeman and spin-orbit couplings needed for the topological applications.
For these purposes, an important merit of such experiments is their high degree of tunability.
For instance, in the experiment of Ref.~\cite{vanwoerkom2016} three separate knobs could be tuned to study the behavior of Andreev bound states: the phase difference $\phi$ across the Josephson junction, the transparency of the junction (controlled by a local gate underneath the weak link), and the magnetic field $B$ (which was applied parallel to the wire).
Thanks to this high tunability, the measurement of the Andreev spectra can allow one to obtain a great wealth of information about the properties of the device.

In order to understand existing experiments and design future ones, it would be beneficial to have a detailed theory of the Andreev bound states, describing their behavior as the magnetic field and other system parameters are continuously varied.
This work aims at providing such theory, focusing on the simple yet experimentally relevant situation of a short, single-channel nanowire junction placed in a magnetic field parallel to the wire (see Fig.~\ref{fig:layout}).
Our theory covers all the important regions of the phase diagram, as depicted in Fig.~\ref{fig:linearization_limits}.
We pay particular attention to the behavior of the Andreev bound states in the topologically trivial phase, since such knowledge may be important to assign experimental data to the correct place of the phase diagram.
Aside from the Andreev energy spectrum, we also study in detail the magnetic field dependence of the matrix elements which determine the absorption strength of microwave-induced Andreev transitions.

The study of Andreev bound states in Josephson junctions with broken time-reversal and/or spin-rotation symmetries is a very rich topic of research, covered by a large and diverse body of previous works \cite{altland1997,bezuglyi2002,kwon2004,krive2004,krive2005,buzdin2005,dellanna2007,beri2008,michelsen2008,reynoso2008,fu2009,mengcheng2012,reynoso2012,sanjose2013,beenakker2013,yokoyama2013,yokoyama2014,heck2014,mironov2015,cayao2015,dolcini2015,konschelle2015,vayrynen2015,konschelle2016,nesterov2016,rasmussen2016,peng2016,sticlet2017}.
In many of the existing studies it is assumed that the time-reversal symmetry breaking is only operative in the weak link, while the effect of the magnetic field in the superconducting parts of the device is disregarded.
At the technical level, this means that the effect of time-reversal symmetry breaking is incorporated in the scattering matrix of the junction but not in the description of the superconducting electrodes.
In the present context, however, it is crucial to include the effect of the magnetic field on the proximity-induced pairing occurring in the nanowire segments which are in direct contact with the superconductors.
Indeed, in experiments aimed at reaching the topological phase, the purpose of the magnetic field is not to influence the local properties of the weak link, but to change the nature of the superconducting pairing induced in the nanowire (whether or not the topological phase is actually reached).
The theory of Andreev bound states developed here, therefore, removes the aforementioned assumption and incorporates the effect of the magnetic field in the entire semiconducting nanowire.

Let us summarize the main results presented in this work, and at the same time outline the layout of the paper.
In Sec.~\ref{sec:model}, we discuss the nanowire model and the different approximations used in this work.
We then derive a determinant equation, Eq.~\eqref{eq:determinant_equation}, which allows us to solve for the discrete part of the spectrum, i.e. to determine the Andreev bound state energies and their wave functions.
The bound state equation \eqref{eq:determinant_equation} makes use of the transfer matrix of the junction, unlike the commonly adopted approach based on the scattering matrix \cite{beenakker1991,beenakker1992}, but akin to previous examples appearing in the literature \cite{shumeiko1993,shumeiko1997,zazunov2005,zazunov2014}.

In a short single-channel junction, the sub-gap spectrum consists of a doublet of Andreev bound states.
In Sec.~\ref{sec:andreev_spectrum}, we study the energies of this Andreev doublet by solving the bound state equation both analytically and numerically
\footnote{The numerical code accompanying this work is available online at
\url{https://github.com/bernardvanheck/andreev_spectra}
}.
The Section begins with a review of basic concepts regarding the excitation spectrum of Josephson junctions (Sec.~\ref{sec:andreev_spectrum_general_discussion}) and of known results about the Andreev bound states at zero magnetic field (Sec.~\ref{sec:SolutionAtZeroField}).
We then discuss the important features of magnetic field dependence of the Andreev bound state energies at both low and high electron density, and in both the topological and trivial phases (Sec.~\ref{sec:ABS_in_field} and Fig.~\ref{fig:abs_spectra}).
In particular, we present analytical results for the effective g-factor which determines the linear energy splitting of the Andreev doublet in a small magnetic field, see Sec.~\ref{sec:g-factor} and specifically Eqs.~\eqref{eq:g_factor_perfect_transmission}-\eqref{eq:g_counterpropagating}.
We show that the g-factor of the Andreev bound states can be strongly suppressed by spin-orbit coupling and/or high electron density.
The resulting g-factor can be much smaller than the g-factor of the conduction electrons in the normal state.
Equations~\eqref{eq:g_factor_perfect_transmission}-\eqref{eq:g_counterpropagating} also indicate that a measurement of the Andreev bound state g-factor, for instance by means of tunneling spectroscopy, can provide information about the other relevant system parameters.
In Sec.~\ref{sec:fermi_level_crossings} we discuss the appearance of Fermi level crossings in the Andreev spectrum.
The presence of Fermi level crossings is significant because it signals a change of the ground state fermion parity of the junction.
These ``fermion parity switches'' can be used as a signature of the topological phase.
Namely, if the nanowire is in the topological (trivial) phase, the number of fermion parity switches occurring as the phase difference $\phi$ is advanced by $2\pi$ must be odd (even).
In the topological phase, this leads to the well-known $4\pi$ periodicity of the phase dependence of the Andreev bound state energies.
The occurrence of fermion parity switches in the trivial phase has so far attracted less attention: here we show that they can appear once the magnetic field crosses a threshold value $B_\textrm{sw}$, which depends sensitively on the transparency of the junction and on the strength of the spin-orbit coupling (see Fig.~\ref{fig:b_sw_vs_spin-orbit}).

In Sec.~\ref{sec:current} we turn our attention to the Josephson current carried by the Andreev bound states, introdu{}cing the current operator and briefly discussing the magnetic field dependence of the current-phase relation (Fig.~\ref{fig:cpr}).
The matrix elements of the current operator between the Andreev bound states determine not only the equilibrium properties of the junction, but also its response to a microwave field.
The microwave irradiation of the junction can induce two types of transitions within the Andreev bound state doublet: both are discussed in Sec.~\ref{sec:absorption} within the linear response regime, appropriate if the applied microwave field is weak.
In the first and most notable type of transition, microwaves resonantly excite a Cooper pair from the superconducting condensate to the Andreev doublet.
In the second type of transition, instead, microwaves excite one quasiparticle from the first to the second Andreev bound state.
The two transitions are distinguished by the parity of the number of quasiparticles involved and so, for brevity, we will refer to them as the ``even'' and ``odd'' transition respectively.
The even transition is present already at zero magnetic field and, being very bright, is the most easily observed in experiment.
The odd transition requires a quasiparticle to be present in the initial state of the junction, either due to a non-equilibrium population or as a consequence of a fermion parity switch.
At zero magnetic field, the odd transition is not observable, since in this case the microwave field cannot induce a transition within the degenerate doublet of Andreev levels.
However, it may become visible in the presence of both Zeeman and spin-orbit couplings.
The magnetic field dependence of the current matrix elements for the even and odd transitions is studied in Secs.~\ref{sec:absorption_even} and \ref{sec:absorption_odd} respectively.
The study reveals that the odd transition, while characterized by a non-zero current matrix element, remains much weaker than the even transition over a wide range of system parameters, including for magnetic fields $B>B_\textrm{sw}$.
An important consequence of this fact is that, at low temperatures, the absorption spectrum of the junction should exhibit a sudden drop in visibility if the junction is driven through a fermion parity switch by varying the magnetic field or the phase difference (see Fig.~\ref{fig:ReY_vs_B_high_mu}).

\section{Model and Andreev bound state equation}
\label{sec:model}

\begin{figure}
\begin{center}
\includegraphics[width=\columnwidth]{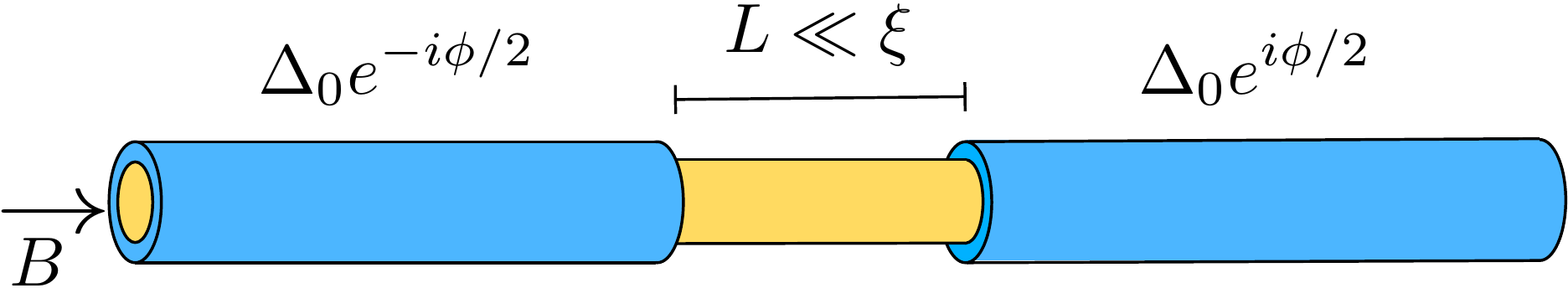}
\caption{The system studied in this work: a Josephson junction with length $L$ made out of a semiconducting nanowire with strong spin-orbit coupling (typically InAs or InSb, yellow) in proximity to a superconductor (typically Al, blue). The proximity effect induces an effective $s$-wave pairing in the nanowire, with gap $\Delta_0$. We focus on the case of a short Josephson junction with $L\ll \xi$, where $\xi$ is the induced coherence length. A parallel magnetic field $B$ and a phase bias $\phi$ are applied to the nanowire. The transparency of the junction can be controlled via a local gate. \label{fig:layout}}
\end{center}
\end{figure}

\begin{figure}
\begin{center}
\includegraphics[width=\columnwidth]{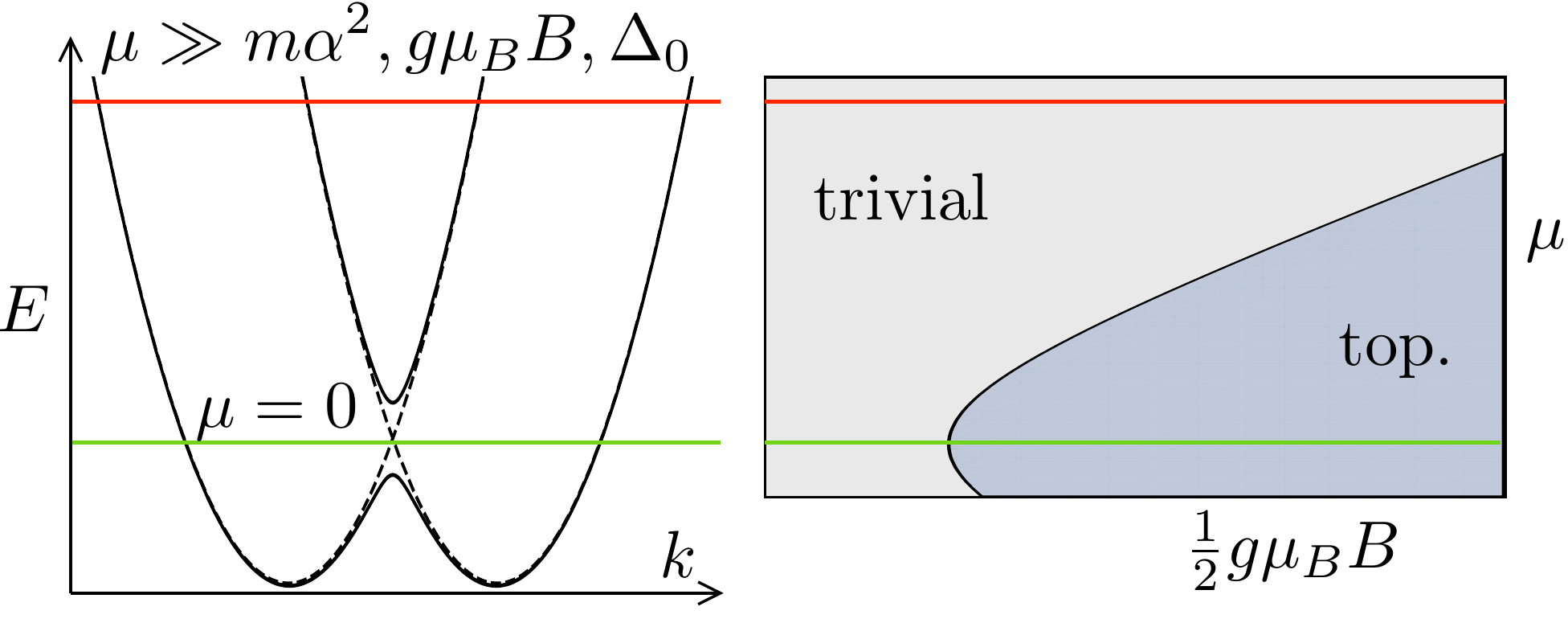}
\caption{Sketch of the nanowire band structure (left) and phase diagram (right). We study the Andreev bound state spectrum by linearizing the spectrum in the two limits $\mu\gg m\alpha^2, \Delta_0, g\mu_B B$ (red line), and $m\alpha^2 \gg \mu, \Delta_0, g\mu_B B$ (green line). In the latter limit the chemical potential can be tuned to be inside the Zeeman gap, and the system can enter the topological phase upon increasing the magnetic field.}
\label{fig:linearization_limits}
\end{center}
\end{figure}

Our investigation is based on the well-studied model of a one-dimensional (1D) quantum wire with Rashba spin-orbit coupling, a Zeeman field applied parallel to the wire, and a proximity-induced $s$-wave pairing \cite{lutchyn2010,oreg2010}.
We consider the Josephson junction geometry shown in Fig.~\ref{fig:layout}.
In the limit $L/\xi\to 0$, we can treat the junction as a point-like defect situated at a position $x=0$.
Furthermore, provided that the length of the entire nanowire is much larger than $\xi$, we can ignore complications arising from its finite size and treat it as an infinite system in the $x$ direction.
The effective BCS Hamiltonian of this system is (we set $\hbar=1$)
\begin{align}\label{eq:wire_model}
H &= \frac{1}{2}\int \de x \,\psi^\dagger(x)\,\left[\left(-\frac{\pd_x^2}{2m}-i\alpha\pd_x\,\sigma_z - \mu\right)\tau_z \right. \\\nonumber
&\left.- \frac{1}{2}\,g\,\mu_B B\,\sigma_x - \Delta_0\,\tau_x\,\e^{-i\phi\,\sgn(x)\,\tau_z/2} + V \delta(x)\,\tau_z\right]\,\psi(x)\,.
\end{align}
In this Hamiltonian, the field operator $\psi(x)$ is the usual four-component Nambu spinor, $\psi=(\psi_\up\,,\,\psi_\down\,,\,\psi^\dagger_\down\,,\,-\psi^\dagger_\up)^T$.
The two sets of Pauli matrices $\sigma_{x,y,z}$ and $\tau_{x,y,z}$ act in spin and Nambu space, respectively.
Furthermore, $m$ is the effective mass, $\alpha$ is the strength of the Rashba spin-orbit coupling, $B$ is the applied magnetic field, $g$ is the effective g-factor, $\mu_B$ is the Bohr magneton, $\Delta_0$ is the proximity-induced pairing gap, and $\phi$ is the gauge-invariant phase difference across the Josephson junction, and $\mu$ is the chemical potential measured from the middle of the Zeeman gap at $k=0$ (see Fig.~\ref{fig:linearization_limits}).
Finally, the phenomenological parameter $V$ is the strength of a point-like scatterer which models a potential barrier; later on $V$ will be related to measurable properties of the junction.
In practice, the scattering term $V\delta(x)$ enters purely in the boundary condition for $\psi(x)$ at the position of the junction,
\begin{subequations}\label{eq:boundary_condition_wire_model}
\begin{align}
\psi(0^+)&=\psi(0^-)\,,\\\label{eq:boundary_condition_derivative}
\pd_x\psi(0^+)-\pd_x\psi(0^-) &= 2 m V\,\psi(0^-)\,.
\end{align}
\end{subequations}

Being a purely one-dimensional effective model, the Hamiltonian in Eq.~\eqref{eq:wire_model} does not incorporate all the complexity of real devices.
For instance, the orbital effect of the magnetic field is not included in our analysis: this is well justified if the cross-section $A$ of the nanowire is small, so that at a given field $B$ the flux piercing the cross-section is much smaller than a flux quantum ($BA\ll h/e$)~\footnote{Nanowires currently studied in experiments have a nominal area of about $(100\;\text{nm})^2$, but the lowest subbands wave functions are most likely extended over a much smaller area $\sim 1000$ nm$^2$ due to gate confinement and band bending.
Thus, they are only sensitive to a portion of the total applied flux.
One may estimate that the orbital effect of the magnetic field can not be dominant for $B\lesssim 1$~T, while the experiment reported in Ref.~\cite{vanwoerkom2016} covered a field range smaller than $500$~mT.}.
The pairing strength $\Delta_0$, the spin-orbit strength $\alpha$ and the g-factor $g$ appearing in Eq.~\eqref{eq:wire_model} should be viewed as phenomenological parameters.
The value of $\Delta_0$, in particular, strongly depends on the transparency  of the semiconductor-superconductor interface.

By virtue of simplicity, the Hamiltonian~\eqref{eq:wire_model} has become a paradigmatic model in the study of Majorana physics in hybrid semiconducting-superconducting system \cite{stanescu2013}.
As is well known, it exhibits two distinct topological phases (see Fig.~\ref{fig:linearization_limits}).
At high chemical potential and/or low magnetic fields, the system is in a trivial superconducting phase with a conventional $2\pi$-periodicity of the ground state energy with respect to the phase difference $\phi$.
At low chemical potential, and provided that the condition $\tfrac{1}{2}g\mu_B B >(\mu^2+\Delta_0^2)^{1/2}$ is satisfied, the system is instead in a topological superconducting phase.
In the geometry of Fig.~\eqref{fig:layout}, the hallmark of the topological phase is the $4\pi$-periodicity of the ground state energy (for a fixed global fermion parity) with respect to $\phi$, which is associated with the presence of two coupled Majorana zero modes at the junction.
The two phases are separated by a critical line $B_c(\mu)=2 (\mu^2+\Delta_0^2)^{1/2}/(g\mu_B)$ at which the energy gap in the nanowire vanishes, marking a topological phase transition.
Note that this criterion is appropriate if the transparency of the semiconductor-superconductor interface is low, so that the coupling between the two materials is weak: in the opposite limit of strong coupling, the critical field $B_c$ may depend only weakly on $\mu$ \cite{sticlet2017}.
A more complicated phase diagram in the $(\mu, B)$ plane emerges in nanowires with multiple transport channels \cite{stanescu2011}, but we do not consider this situation in this paper.

For both regions of the phase diagram, we are interested in the Andreev level spectrum.
That is, we want to find the discrete spectrum of sub-gap states with energy $E<\Delta(B)$, which are localized at the junction via the mechanism of Andreev reflection at the two superconducting interfaces.
The energy $\Delta(B)$ is the spectral gap of the continuous part of the spectrum [at zero field, $\Delta(0)\equiv \Delta_0$].
In what follows, we will often omit the field argument, $\Delta(B)\equiv \Delta$.
In the short junction limit, one expects the number of sub-gap states to be less than or equal to the number of pairs of left/right propagating modes at the Fermi level in the normal state of the nanowire.
Thus, the discrete spectrum of the Hamiltonian in Eq.~\eqref{eq:wire_model} should consist of at most either one or two Andreev levels, depending on whether the system is in the topological or in the trivial phase.

An established way to compute the Andreev level spectrum is the scattering approach \cite{beenakker1991,beenakker1992,lesovik2011,heck2014,sticlet2017}.
In this approach, the usual wave function matching problem for bound states is cast in terms of a scattering matrix $S_N(E)$ which characterizes the junction, and a second scattering matrix $S_A(E)$ which describes the Andreev reflection from the superconducting leads.
The two matrices can be combined into a determinant equation for the bound state energies, $\det[1-S_A(E)S_N(E)]=0$.
This approach is particularly advantageous if the following conditions are satisfied.
First, the effect of a magnetic field can be neglected in the superconducting leads, so that the gap and the matrix structure of $S_A$ are independent of magnetic field.
Second, the normal reflection at the superconducting interface can be neglected -- this is the so-called Andreev approximation \cite{andreev1964}:
it requires $\Delta_0\ll E_F$ (where $E_F$ is the Fermi energy measured from the bottom of the conduction band) and amounts to linearizing the electron dispersion in the normal state.
Third, the junction is short, so that the energy dependence of $S_N(E)$ can be neglected as long as $E<\Delta_0$.
When combined, the first and second conditions guarantee that $S_A(E)$ is a simple sparse matrix whose energy dependence enters only as a prefactor, $S_A(E)\propto \exp[i \arccos(E/\Delta_0)]$.
This circumstance greatly simplifies the solution of the problem, as exemplified by the fact that the determinant equation can be transformed into a finite-dimensional linear eigenvalue problem for $E$ \cite{heck2014}.

However, as mentioned in the introduction, for our purposes it is crucial to include the effect of magnetic field in the entire system, rather than in the junction alone.
The motivation for doing so is two-fold.
To begin with, many recent experiments focused on InAs nanowires with epitaxial Al: in this geometry, a parallel field penetrates uniformly the thin aluminum shell.
Furthermore, the study of the evolution of the Andreev levels in the different regions of the phase diagram --- and in particular across the topological transition --- requires that we solve for the Andreev energies taking into account the magnetic field dependence of the spectral gap of the continuous spectrum.
Unfortunately, once a magnetic field is included, the Andreev reflection amplitude is not unique anymore but may depend on the initial and final spin and/or orbital states.
As a consequence, the calculation of $S_A(E)$ becomes non-trivial and strongly dependent on the particular Hamiltonian describing the leads.
To overcome this complication, we take a slightly different route and derive an equivalent bound state equation for the Andreev spectrum which generalizes to the more complicated cases in a transparent fashion.

The first step in the derivation is the linearization of the model in Eq.~\eqref{eq:wire_model}, which we perform in two different limits allowing us to cover all relevant regimes of the phase diagram (see Fig.~\ref{fig:linearization_limits}).
The first limit is that of the high density, $\mu\gg m\alpha^2, \Delta_0, g\mu_B B$.
For such high values of the chemical potential the nanowire will not enter the topological phase in a realistic range of magnetic fields, thus we will use this limit to model a topologically trivial nanowire.
The second limit is that of low density, when the Fermi level is inside the Zeeman gap .
This is the ``helical'' regime of the Rashba nanowire: in the normal state, the low-energy theory contains only a pair of counter-propagating modes at finite wave vectors, as well as a gapped pair of modes close to $k=0$.
The line $\mu=0$ in the phase diagram coincides with the optimal point at which the critical field is minimal, $B_c=\Delta_0$ (see Fig.~\ref{fig:linearization_limits}), so this limit will allow us to study the Andreev spectrum in the topological phase and around the phase transition.
In both limits we will require the Andreev approximation to hold.
The Andreev approximation is automatically satisfied in the high density regime, when $\Delta_0\ll\mu$.
In the low density regime, the chemical potential is low and the Fermi energy is set by the spin-orbit energy, $E_F\sim m\alpha^2$.
Thus, in this limit we must assume the spin-orbit energy to be the dominating energy scale: $m\alpha^2\gg\Delta_0, g\mu_B B, \mu$.

In the two following subsections, we carry out the linearization procedure in these two limits, which will then allow us to derive the bound state equation that we seek.

\subsection{Linearization for $\mu\gg \Delta_0, g\mu_B, m\alpha^2$}
\label{subsec:linearization_mu}

In the limit of a high chemical potential, we may linearize the normal state dispersion around the Fermi wave vectors $\pm k_F=\pm(2m \mu)^{1/2}$.
That is, we write the field $\psi(x)$ as a linear superposition of left- and right-moving components,
\begin{equation}\label{eq:linearized_field_kF}
\psi(x) = \e^{-ik_F x}\,\psi_L(x) + \e^{ik_Fx}\,\psi_R(x)\,.
\end{equation}
Since we are interested in the energy spectrum in a range of energies of order $\Delta$ around the Fermi level, we can assume that $\psi_L(x)$ and $\psi_R(x)$ vary over length scales much larger than $k_F^{-1}$.
We may therefore use Eq.~\eqref{eq:linearized_field_kF} in the Hamiltonian \eqref{eq:wire_model}, organize the resulting expression as an expansion in powers of $k_F^{-1}$, and only keep the largest terms.
The last step also includes neglecting quickly oscillating terms $\propto \e^{\pm ik_F x}$.
The result of this procedure can be concisely presented by introducing an eight-component field vector $\Psi = (\psi_R\,,\,\psi_L)^T$.
In terms of the slowly-varying field $\Psi(x)$, the low-energy Hamiltonian of the nanowire is
\begin{align}\nonumber
H \approx \frac{1}{2}&\int \de x \,\Psi^\dagger(x)\,\left[-i v_F\,\tau_z\,s_z\,\pd_x + \alpha k_F\,\tau_z\,s_z\,\sigma_z \right.\\
&\left.- \tfrac{1}{2}\,g\,\mu_B B\,\sigma_x - \Delta_0\,\tau_x\,\e^{-i\phi\,\sgn(x)\,\tau_z/2}\right]\,\Psi(x)\,.\label{eq:linearized_hamiltonian_high_mu}
\end{align}
with $v_F=k_F/m$. Here, we have introduced a new set of Pauli matrices $s_{x,y,z}$ which act in the space of left- and right-movers. Let us now describe the low-energy modes described in this linearized Hamiltonian.

\begin{figure}
\begin{center}
\includegraphics[width=\columnwidth]{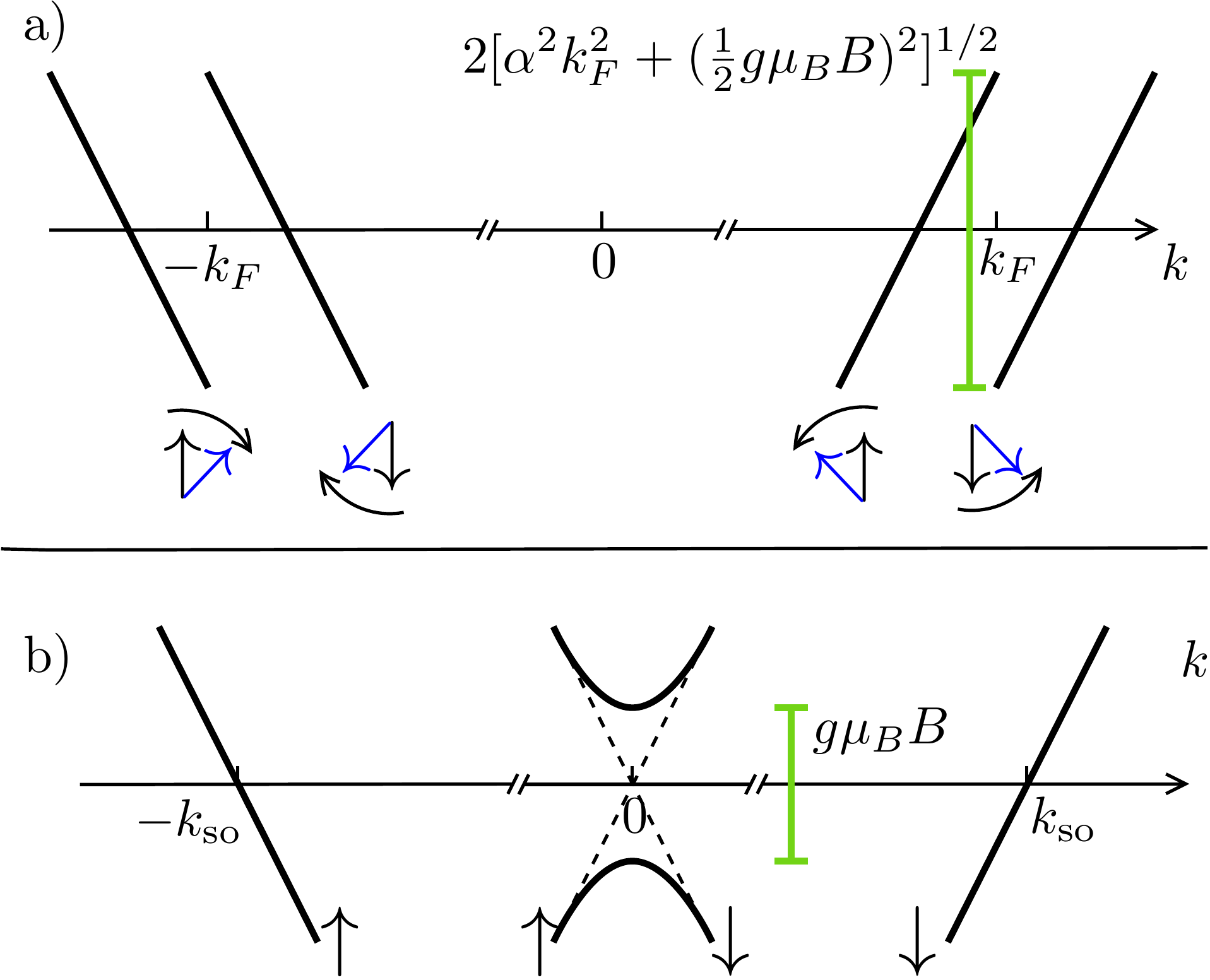}
\caption{Sketch of the dispersion in the normal state after linearization. Panel (a): dispersion for $\mu\gg g\mu_B B, m\alpha^2$. The black (blue) arrows denote the spin direction for each branch of the spectrum at zero (finite) magnetic field $B$. Panel (b): dispersion for $\mu=0$. Here $k_\textrm{so}=2m\alpha$. The magnetic field gaps out the crossing at $k=0$, which is between states with opposite spins. \label{fig:linearized_dispersions}}
\end{center}
\end{figure}

As illustrated in Fig.~\ref{fig:linearized_dispersions}a, around each Fermi point there are two branches in the spectrum of the normal state.
The two branches are separated in energy by an amount $2 [\alpha^2k_F^2 + (\tfrac{1}{2}g\mu_B B)^2]^{1/2}$ due to the combined effect of spin-orbit and Zeeman coupling.
At finite $B$, the spin of each propagating mode is rotated with respect to its orientation at $B=0$ (see arrows in Fig.~\ref{fig:linearized_dispersions}a).
The rotation angle is
\begin{equation}
\theta =  \arccos\,\frac{\alpha k_F}{[\alpha^2k_F^2 + (\tfrac{1}{2}g\mu_B B)^2]^{1/2}}\,.
\end{equation}
and the rotation plane is defined by the Rashba and Zeeman fields.
The spin rotation is clockwise (counter-clockwise) for left (right) movers and it can be incorporated in the definition of the field $\Psi$ via a unitary transformation $S$ (see for instance Ref.~\cite{nesterov2016}),
\begin{equation}
\Psi_S(x) = S\,\Psi(x)\;,\;\;S=\exp\,[-i\,(\theta/2)\,\tau_zs_z\sigma_y]
\end{equation}
This rotated basis diagonalizes the homogeneous Hamiltonian of the wire in the normal state.
When we express the Hamiltonian in terms of the rotated field $\Psi_S$, we find
\begin{align}\nonumber
H &\approx \frac{1}{2}\int \de x \,\Psi_S^\dagger(x)\,\left[-i v_F\tau_zs_z\,\pd_x + \alpha k_F\sec\theta\,\tau_z\,s_z\,\sigma_z \right.\\
&\left.- \Delta_0\,(\cos\theta\,\tau_x + \sin\theta\,\tau_ys_z\sigma_y)\,\e^{-i\phi\,\sgn(x)\,\tau_z/2}\right]\,\Psi_S(x)\,.
\end{align}
This form of the Hamiltonian reveals how the tilting of the modes' spin affects the pairing.
At $B=0$, the $s$-wave pairing does not mix the inner and outer branches of the spectrum since it requires the spins of the two paired electrons to be anti-parallel.
At finite field, however, a pairing coupling with strength $\Delta_0\sin\theta\approx \Delta_0 \,(g\mu_B B)/(\alpha k_F)$ emerges between the inner and outer branches, due to the fact that the spin tilts in opposite directions for left and right movers.

To complete the linearization procedure, we must provide the boundary conditions for the field $\Psi$ which are due to the scattering term $V \delta(x)\,\tau_z$ in the original model of Eq.~\eqref{eq:wire_model}.
The boundary conditions for $\Psi$ can be derived by using Eq.~\eqref{eq:linearized_field_kF} in Eqs.~\eqref{eq:boundary_condition_wire_model} and neglecting terms $\propto \pd_x\psi_{L,R}$ with respect to terms $\propto k_F$.
The resulting boundary conditions can be arranged in the following form,
\begin{equation}\label{eq:boundary_condition_Psi}
\Psi(0^+) = T\,\Psi(0^-)\,,
\end{equation}
with
\begin{equation}\label{eq:transfer_matrix}
T = 1 - i (V/ v_F)\,s_z + (V/v_F)\,s_y\,.
\end{equation}
The matrix $T$ is, in fact, the transfer matrix associated with the point-like scatterer $V\delta(x)$ in the original model, computed at the Fermi level.
The term $\propto s_y$ is a backscattering term, while the term $\propto s_z$ corresponds to forward scattering.
The transmission probability $\tau$ through the junction in the normal state is related to the dimensionless parameter $V/v_F$,
\begin{equation}
\tau = \frac{1}{1+(V/v_F)^2}\,.
\end{equation}
The transfer matrix obeys a ``pseudo-unitarity'' property,
\begin{equation}\label{eq:pseudounitary_transfer_matrix}
T^\dagger = s_z\,T^{-1}\,s_z\,,
\end{equation}
which is the equivalent of the most universally known unitarity of the scattering matrix.

For the rotated field $\Psi_S$, we must use a rotated transfer matrix $T_S=STS^\dagger$,
\begin{align}\nonumber
T_S &= 1 - i (V/ v_F)\,s_z+ (V/v_F)\,\cos\theta\,s_y\\
&-(V/v_F)\,\sin\theta\,\tau_zs_z\sigma_y\,.
\end{align}
We see that, similar to the pairing, the backscattering terms are changed when projected to the basis of momentum eigenstates of the nanowire.
At zero field, only a single backscattering channel is open for each mode, because scattering preserves spin.
At a finite field $B$, two backscattering terms appear, due to the fact that the spin of each left-moving mode has non-zero projection on the spin of both right-moving modes.

\subsection{Linearization for $m\alpha^2 \gg \Delta_0, g\mu_B B, \mu$}
\label{subsec:linearization_mu=0}

When the chemical potential is low, the linearization procedure must take into account that the position of the Fermi points strongly depends on the spin orientation, since the Fermi points are shifted by the Rashba spin-orbit coupling.
Specifically, the Fermi points for modes with spin down (up) are situated at $k=2 m\alpha$ ($k=0$) for right-movers and at $k=0$ ($k=-2m\alpha$) for left-movers; see Fig.~\ref{fig:linearized_dispersions}b.
The linearization of the model therefore begins by writing the field in the following form \cite{klinovaja2012},
\begin{equation}\label{eq:linearized_field_mu=0}
\psi(x) = \e^{-im\alpha x\,(1+\sigma_z)}\,\psi_L(x) + \e^{im\alpha x\, (1-\sigma_z)}\,\psi_R(x)\,.
\end{equation}
Note the presence of the spin-dependent factors in the exponentials, which take into account the dependence of the Fermi points on spin.
From here, we proceed as in the previous subsection: assuming that the left- and right- moving fields $\psi_L(x)$ and $\psi_R(x)$ vary over length scales much larger than $(m\alpha)^{-1}$, we replace Eq.~\eqref{eq:linearized_field_mu=0} in Eq.~\eqref{eq:wire_model} with $\mu=0$, and neglect all quickly oscillating terms $\propto \e^{\pm 2im\alpha x}$.
Note that, in doing this, it is essential to assume that the spin-orbit energy dominates over the other energy scales.
In other words, the spin-orbit length $(m\alpha)^{-1}$ takes the role of the Fermi wavelength as the microscopic length scale of the model.

As a result we obtain the following linearized Hamiltonian of freely propagating modes,
\begin{align}\nonumber
H \approx \frac{1}{2}\int \de x \,&\Psi^\dagger(x)\,\left[-i \alpha\tau_zs_z\,\pd_x - \tfrac{1}{4}g\mu_B B(s_x \sigma_x - s_y \sigma_y) \right.\\
&\left.- \mu\tau_z - \Delta_0\,\tau_x \,\e^{-i\phi\,\sgn(x)\,\tau_z/2}\right]\,\Psi(x)\,.\label{eq:linearized_hamiltonian_mu=0}
\end{align}
Here, as in Eq.~\eqref{eq:linearized_hamiltonian_high_mu}, $\Psi=(\psi_R, \psi_L)^T$ and the set of Pauli matrices $s_{x,y,z}$ acts in the grading of left- and right-movers.

When written in terms of the components of the vector $\Psi$, the Zeeman term in Eq.~\eqref{eq:linearized_hamiltonian_mu=0} is proportional to $\psi^\dagger_{R\up} \psi_{L\down}$.
It is a mass term which gaps out the two counter-propagating modes with opposite spin crossing at $k=0$ (see Fig.~\ref{fig:linearized_dispersions}).
Note that, once this Zeeman gap is formed at the Fermi level, the presence of a scattering impurity may lead to Fano resonances \cite{cayao2015}.
The Fano resonances are due to the formation --- in the normal state --- of quasi-bound states with a characteristic decay length $\alpha/g\mu_B B$.
The quasi-bound states originate from the inverted part of the parabolic spectrum close to $k=0$, and in principle they can lead to a strong dependence of the transmission of the junction in the normal state on energy \cite{cayao2015,nesterov2016}.
We may neglect complications associated with their presence by assuming that the junction is short enough so that $L\ll \alpha/g\mu_B B$.
With this assumption, boundary conditions for $\Psi$ can also be derived as in the previous subsection.
We obtain the same transfer matrix $T$ of Eq.~\eqref{eq:transfer_matrix}, except with the velocity $v_F$ replaced by $\alpha$; the transmission probability is now $\tau=1/(1+V^2/\alpha^2)$.

\subsection{Bogoliubov-de Gennes equations and bound state determinant condition}

At this point, in either of the two limits $\mu\gg \Delta_0$ and $\mu=0$, our task is reduced to the solution of a system of Bogoliubov-de Gennes (BdG) equations
\begin{align}\label{eq:BdG_equations}
\left[-iv\tau_zs_z\pd_x+O_N-\Delta_0\tau_x\e^{-i\phi\sgn(x)\tau_z/2}\right]\Phi = E\Phi\,,
\end{align}
for an eight-component Nambu wave function $\Phi(x)$ \footnote{Note that we use the letter $\Phi$ for the BdG wave functions, and $\Psi$ for the corresponding second-quantized fields}, to be solved with the boundary condition $\Phi(0^+)=T\Phi(0^-)$.
The pseudo-unitarity of the transfer matrix $T$, Eq.~\eqref{eq:pseudounitary_transfer_matrix}, guarantees that the kinetic energy in the BdG equations remains a Hermitian operator when removing the point $x=0$ from its domain.
In Eq.~\eqref{eq:BdG_equations}, $O_N=\alpha k_F \tau_z s_z \sigma_z - \tfrac{1}{2} g\mu_B \sigma_x$ for $\mu\gg\Delta_0, g\mu_B B, m\alpha^2$ while $O_N =- \tfrac{1}{2}g\mu_B B \,\tfrac{1}{2}(s_x \sigma_x - s_y \sigma_y)-\mu\tau_z$ for $m\alpha^2\gg \Delta_0, g\mu_B B, \mu$.
The velocity $v$ is a placeholder for $v_F$ in the former case, and for $\alpha$ in the latter.

As is well known \cite{degennes}, the BdG equations are inherently equipped with a particle-hole symmetry represented by an anti-unitary operator $\mathcal{P}$.
The particle-hole symmetry dictates that for each solution $\Phi$ of Eq.~\eqref{eq:BdG_equations} at energy $E$ there must be an orthogonal solution $\mathcal{P}\Phi$ at energy $-E$.
In our case, $\mathcal{P}=\tau_ys_x\sigma_y\,\mathcal{K}$, with $\mathcal{K}$ the complex conjugation operator.
The presence of particle-hole symmetry, and the corresponding doubling of the spectrum, is a consequence of the unphysical doubling of the Hilbert space coming from the introduction of Nambu indices; more fundamentally, it is a consequence of the mean-field approximation which allowed us to express the Hamiltonian \eqref{eq:wire_model} as a quadratic form of $\psi$ and $\psi^\dagger$ \cite{altland1997}. 

Once the complete spectrum $\{\pm\, E_n\}$ of the BdG equations is known, the field operator $\Psi(x)$ can be written in the eigenmode expansion
\begin{equation}\label{eq:eigenmode_expansion}
\Psi(x) = \sum_n\,\Gamma_n\,\Phi_n(x) + \Gamma^\dagger_n\,[\mathcal{P}\Phi_n(x)]\,.
\end{equation}
Here, $\Gamma_n$ and $\Gamma_n^\dagger$ are Bogoliubov annihilation and creation operators, obeying fermionic anticommutation relations.
They diagonalize the Hamiltonian,
\begin{equation}
H = \sum_n E_n\left(\Gamma^\dagger_n\Gamma_n - \tfrac{1}{2}\right)\,.
\end{equation}
Our goal is to find the bound state solutions of Eq.~\eqref{eq:BdG_equations}, which have $\abs{E}<\Delta$.
In order to do so, we first bring the BdG equations to a more convenient form by a change of variable, 
\begin{equation}
\Phi(x) =\e^{i \phi\sgn(x)\tau_z/4}\,\tilde{\Phi}(x)\,.
\end{equation}
The role of this transformation is to make the spatial dependence of the superconducting phase more easily tractable.
The wave function $\tilde\Phi(x)$ satisfies a modified boundary condition at the origin,
\begin{equation}\label{eq:boundary_condition}
\tilde\Phi(0^+) = \e^{-i\phi \tau_z/2}\,T\,\tilde{\Phi}(0^-)\,.
\end{equation}
Next, we define the Green's function $G(x, E)$ by
\begin{equation}
[E-H_\textrm{BdG}(\pd_x)]\,G(x, E) = i v \tau_z s_z \delta(x)\,,
\end{equation}
where the operator $H_\textrm{BdG}(\pd_x)$ is the linearized BdG Hamiltonian of the translationally invariant superconducting wire,
\begin{equation}\label{eq:HBdG_homogeneous}
H_\textrm{BdG}(\pd_x) = -iv\tau_zs_z\pd_x +O_N-\Delta_0\,\tau_x\,.
\end{equation}
Note that by definition $G(0^+, E)-G(0^-, E)=1$. Now, using the boundary condition \eqref{eq:boundary_condition}, we may write
\begin{equation}\label{eq:wavefunction}
\tilde\Phi(x) = G(x, E)\,M\,\tilde\Phi(0^-)\,,
\end{equation}
with
\begin{equation}
M = \left(\e^{-i\phi\tau_z/2}\,T-1\right)\,.
\end{equation}
Equation~\eqref{eq:wavefunction} holds for any $x\neq 0$; the wave function is discontinuous at $x=0$.
The Green's function can be computed as
\begin{equation}\label{eq:Green_function}
G(x, E) = -\int_{-\infty}^\infty\,\frac{v\de q}{2\pi i}\frac{\e^{iq x}}{E-H_\textrm{BdG}(q)}\,\tau_z s_z,
\end{equation}
with $H_\textrm{BdG}(q)$ the Fourier transform of Eq.~\eqref{eq:HBdG_homogeneous}. When $\abs{E}<\Delta(B)$, the poles in the integrand of $G(x, E)$ lie away from the real axis and Eq.~\eqref{eq:Green_function} can be computed via a contour integral closing on the upper (lower) half of the complex plane for $x>0$ ($x<0$). Requiring that a non-trivial solution $\tilde\Phi(0^-)$ exists, we obtain from Eq.~\eqref{eq:wavefunction} the following determinant equation for the bound state spectrum:
\begin{equation}\label{eq:determinant_equation}
\det\,\left[1 - G(0^-, E)\,M\,\right]=0\,.
\end{equation}
This bound state equation for the Andreev levels is cast in terms of a transfer matrix $T$ and a Green's function $G(0^-, E)$ for the superconducting leads, rather than in terms of scattering matrices.
As a consequence of the short junction limit considered in this work, the energy dependence of Eq.~\eqref{eq:determinant_equation} is entirely contained in $G(0^-, E)$, while $T$ is independent of energy.
Furthermore, as mentioned at the end of Section~\ref{subsec:linearization_mu=0}, the matrix $T$ contained in Eq.~\eqref{eq:determinant_equation} is the same in both linearization limits when expressed in terms of the transmission probability $\tau$:
\begin{equation}\label{eq:transfer_matrix_with_tau}
T = 1 - i \sqrt{\frac{1-\tau}{\tau}} s_z +\sqrt{\frac{1-\tau}{\tau}} s_y 
\end{equation}
Thus, the differences in the subgap spectrum between the two regimes all arise from $G(x, E)$.
In the following, we compute $G(x, E)$ for the two regimes of interest.
In doing so, we also derive the magnetic field dependence $\Delta(B)$ of the continuum gap.

\subsection{Green's functions and magnetic field dependence of the induced gap}

\subsubsection{Green's function for  $\mu\gg m\alpha^2, g\mu_B B, \Delta_0$}

\begin{figure}
\begin{center}
\includegraphics[width=\columnwidth]{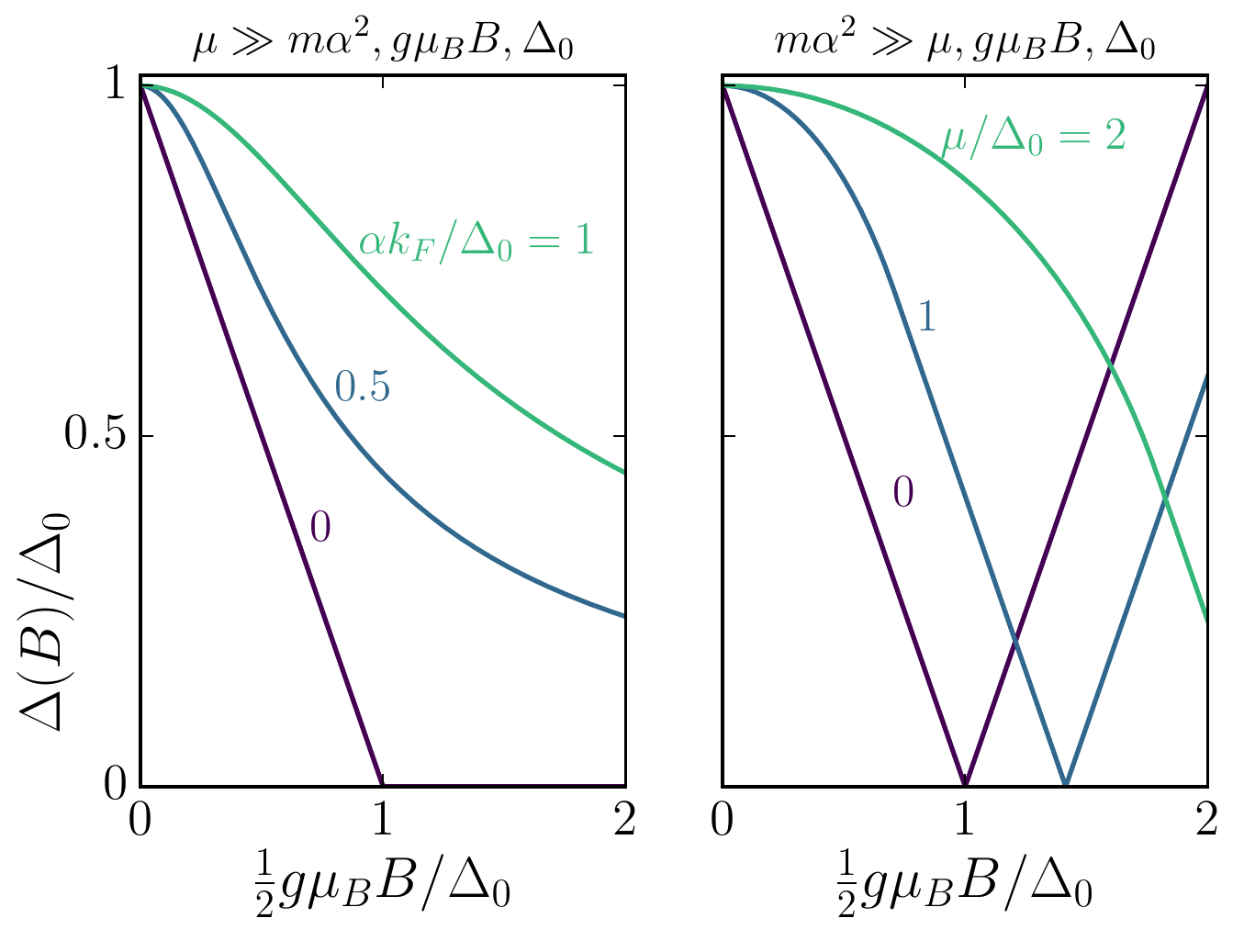}
\caption{Left panel: magnetic field dependence of the proximity-induced gap $\Delta(B)$ in the high chemical potential regime, computed from Eq.~\eqref{eq:gap_high_mu} for different values of the spin-orbit strength, measured by increasing ratios $\alpha k_F/\Delta_0$. When $\alpha=0$, the gap closes for strong enough Zeeman energies, and proximity-induced superconductivity is destroyed. As long as some $s$-wave pairing is induced in the wire, any finite value of spin-orbit strength will prevent such transition to a gapless state to take place, because the spin-orbit coupling prevents a complete alignment of the electron spins with the magnetic field. The spin-orbit coupling also changes the small field behavior of $\Delta(B)$ from linear [$\Delta(B)-\Delta_0\propto B$] to quadratic [$\Delta(B)-\Delta_0\propto B^2$]. For weak spin-orbit coupling strengths, there is still an intermediate range of magnetic fields for which $\Delta(B)$ decreases linearly with field. Right panel: magnetic field dependence of the proximity induced gap $\Delta(B)$ in the low chemical potential regime, computed from Eq.~\eqref{eq:gap_lowmu} for different values of $\mu$. The gap closes at the critical field $B_c(\mu)=2\sqrt{\Delta_0^2+\mu^2}/g\mu_B$. \label{fig:continuum_gap}}
\end{center}
\end{figure}

In order to obtain $G(x, E)$, we must first invert the $8\times 8$ matrix
\begin{align}\nonumber
E-H_\textrm{BdG}(q) &= E - v_Fq \tau_zs_z - \alpha k_F \tau_z s_z \sigma_z \\
&+\tfrac{1}{2} g\mu_B B \sigma_x + \Delta_0 \tau_x.
\end{align}
This task is simplified by the fact that $E-H_\textrm{BdG}(q)$ is a real matrix and thus its inverse must also be real. The result is
\begin{equation}\label{eq:H_inverse}
\frac{1}{E-H_\textrm{BdG}(q)} = \frac{A_0 + A_1\, v_Fq + A_2\,(v_Fq)^2 -\tau_z s_z\,(v_F q)^3}{v_F^4 (q^2-q_0^2)(q^2-q_1^2)}\,.
\end{equation}
Here, $A_0, A_1, A_2$ are $8\times 8$ matrices which do not depend on $q$. Their detailed expressions are:
\begin{widetext}
\begin{subequations}
\begin{align}\nonumber
A_0 &= -(E + \alpha k_F\,\tau_z s_z\sigma_z)\left[\Delta_0^2-E^2+(\alpha k_F)^2 + (\tfrac{1}{2}g\mu_BB)^2\right] + \tfrac{1}{2}g\mu_BB \left[(\alpha k_F)^2 + (\tfrac{1}{2}g\mu_BB)^2-\Delta_0^2 - E^2\right] \sigma_x\\
&+\Delta_0\,\left[\Delta_0^2 - E^2 + (\alpha k_F)^2 - (\tfrac{1}{2}g\mu_BB)^2\right]\tau_x + g\mu_B B\,\Delta_0\,\left(E\,\tau_x\sigma_x-\alpha k_F\,\tau_y s_z \sigma_y\right)\,,\\
A_1 &= \left[(\alpha k_F)^2 + (\tfrac{1}{2}g\mu_BB)^2 + E^2 -\Delta_0^2\right]\,\tau_zs_z + 2E\,\left(\alpha k_F\,\sigma_z - \tfrac{1}{2}g\mu_BB\,\tau_z s_z \sigma_x\right) - 2 \Delta_0\, \alpha k_F\,\tau_x\,\sigma_z\,,\\
A_2 &= \alpha k_F \tau_z s_z \sigma_z - \tfrac{1}{2}g\mu_BB\,\sigma_x + \Delta_0 \tau_x - E\,.
\end{align}
\end{subequations}
There are four simple poles $\pm q_0, \pm q_1$ appearing on the right side of Eq.~\eqref{eq:H_inverse}, given by
\begin{align}\label{eq:poles}
v_F^2\,q_{0,1}^2 &= E^2-\Delta_0^2 + (\alpha k_F)^2 + (\tfrac{1}{2}g\mu_BB)^2 \pm 2i\,\sqrt{(\alpha k_F)^2 (\Delta_0^2 - E^2)-(\tfrac{1}{2}g\mu_BB)^2 E^2}\,.
\end{align}
In order to complete the calculation of the Green's function, we must insert Eq.~\eqref{eq:H_inverse} in Eq.~\eqref{eq:Green_function} and perform the integral over $q$.
Let us choose $q_0$ and $q_1$ to be the two poles with negative imaginary part. Then, using the residue theorem and some simple algebra, we obtain the following expression for the Green's function:
\begin{equation}\label{eq:Ghighmu}
G(x,E) = \frac{1}{2}\frac{1}{v_F^2(q_0^2-q_1^2)}\,\sum_{n=0,1}(-1)^n \frac{\e^{-iq_n \abs{x}}}{v_F q_n}\,\left[A_0 - \sgn(x)\,A_1\,v_F q_n + A_2 (v_Fq_n)^2 + \sgn(x)\,\tau_z s_z\,(v_F q_n)^3\right]\,\tau_z s_z.
\end{equation}
From Eq.~\eqref{eq:poles} we can easily extract the magnetic field dependence of the continuum gap. The gap $\Delta(B)$ is determined by the smallest value of $E$ such that the poles $q_{0,1}$ have zero imaginary part. A few lines of algebra give the following answer:
\begin{equation}\label{eq:gap_high_mu}
\Delta(B) = 
\begin{cases}
\dfrac{\Delta_0\,\alpha k_F}{\left[(\alpha k_F)^2 + (\tfrac{1}{2}g\mu_BB)^2\right]^{1/2}} & \textrm{if}\quad\sqrt{\tfrac{1}{2}g\mu_BB(\Delta_0-\tfrac{1}{2}g\mu_BB)}<\alpha k_F \quad\textrm{or}\quad\tfrac{1}{2}g\mu_BB>\Delta_0\,,\\
\\
\left[(\alpha k_F)^2 + (\Delta_0-\tfrac{1}{2}g\mu_BB)^2\right]^{1/2}& \textrm{if}\quad\sqrt{\tfrac{1}{2}g\mu_BB(\Delta_0-\tfrac{1}{2}g\mu_BB)}>\alpha k_F\,.
\end{cases}
\end{equation}

The behavior of $\Delta(B)$ is discussed in detail in Fig.~\ref{fig:continuum_gap}. Here we only note that $\Delta(B)$ is a smooth function of $B$, and never reaches zero provided that spin-orbit is present (so that $\alpha\neq 0$). These results are true if we assume no suppression of the gap in the parent superconductor which induces the proximity effect in the nanowire.
In the case of InAs nanowires with epitaxial Al, this is justified by the smallness of Al g-factor and shell thickness.

\subsubsection{Green's function for $m\alpha^2\gg g \mu_B, \Delta_0,
\mu$}

At low chemical potentials, we must repeat the same calculation but starting from the BdG Hamiltonian contained in Eq.~\eqref{eq:linearized_hamiltonian_mu=0}. We must first invert the matrix
\begin{equation}\label{eq:E-H_mu=0}
E- H_\textrm{BdG}(q)=E - \alpha q\,\tau_zs_z + \tfrac{1}{4}g\mu_B B\,(s_x \sigma_x - s_y \sigma_y) + \mu\tau_z + \Delta_0\,\tau_x\,.
\end{equation}
In this case, we may simplify the calculation by noting the presence of the unitary symmetry $[E-H_\textrm{BdG}(q), s_z\sigma_z]=0$.
This symmetry is a consequence of the fact that the inner ($k\approx 0$, $s_z\sigma_z=1$) and outer ($k\approx \pm 2 m\alpha$, $s_z\sigma_z=-1$) branches of the linearized spectrum are decoupled in the homogeneous wire (although they are coupled by scattering at the junction).
Furthermore, the outer branches are not coupled to the magnetic field in the linearized Hamiltonian of Eq.~\eqref{eq:linearized_hamiltonian_mu=0}, and so for these modes $s_z$ and $\sigma_z$ are also separately conserved operators.
These facts allow use to separate the inverse of Eq.~\eqref{eq:E-H_mu=0} as a sum of two parts,
\begin{align}\nonumber
\frac{1}{E- H_\textrm{BdG}(q)}&=\frac{1-s_z\sigma_z}{2}\,\left[\frac{1+s_z}{2}\,\frac{(\mu-\alpha q) \tau_z s_z + \Delta_0\tau_x-E}{\Delta_0^2-E^2+(\mu-\alpha q)^2}+\,\frac{1-s_z}{2}\,\frac{(\mu+\alpha q) \tau_z s_z + \Delta_0\tau_x-E}{\Delta_0^2-E^2+(\mu+\alpha q)^2}\right]\\
&+\frac{1+s_z\sigma_z}{2}\,\frac{B_0 + B_1\, \alpha q + B_2\,(\alpha q)^2 - \tau_z s_z\,(\alpha q)^3}{\alpha^4 (q^2-q_0^2)(q^2-q_1^2)}
\label{eq:inverse_E-H_mu=0}
\end{align}
This time, the poles $q_0, q_1$ appearing in Eq.~\eqref{eq:inverse_E-H_mu=0} are given by
\begin{align}\label{eq:poles_mu=0}
\alpha^2 q_{0,1} & = E^2+\mu^2-(\tfrac{1}{2}g\mu_B B)^2-\Delta_0^2\pm 2 i \sqrt{\mu^2(\Delta_0^2-E^2)-\Delta_0^2 (\tfrac{1}{2}g\mu_B B)^2}\,,
\end{align}
while the matrices $B_0, B_1, B_2$ are
\begin{subequations}
\begin{align}\nonumber
B_0 & = E\,\left[E^2-\Delta_0^2-(\tfrac{1}{2}g\mu_B B)^2\right] - \tfrac{1}{2}g\mu_B B\left[\Delta_0^2 + E^2 -(\tfrac{1}{2}g\mu_B B)^2\right]\,s_x\sigma_x + \Delta_0\left[\Delta_0^2 - E^2 -(\tfrac{1}{2}g\mu_B B)^2\right]\,\tau_x\\
& + g\mu_B B\, \Delta_0\,E\,\tau_x s_x \sigma_x\\
B_1 & = \left[E^2 -\Delta_0^2-(\tfrac{1}{2}g\mu_B B)^2\right]\tau_zs_z - g\mu_B B \Delta_0 \tau_y s_x \sigma_y\,,\\
B_2 & = - E + \tfrac{1}{2}g\mu_B B\,s_x \sigma_x + \Delta_0 \tau_x\,.
\end{align}
\end{subequations}
From these expressions, we may compute the Green's function in this regime:
\begin{align}\nonumber
&G(x,E) = \frac{1-s_z\sigma_z}{2}\,\frac{i\,\e^{-i\mu x s_z/\alpha}\e^{-\sqrt{\Delta_0^2-E^2}\abs{x}/\alpha}}{2\sqrt{\Delta_0^2-E^2}}\,\left[\Delta_0\tau_x - E -i \sgn(x)\,\tau_z\,s_z\,\sqrt{\Delta_0^2-E^2}\right]\,\tau_z\,s_z\\
&+\frac{1+s_z\sigma_z}{2}\frac{1}{2\alpha^2(q_0^2-q_1^2)}\sum_{n=0,1}(-1)^n \frac{\e^{-iq_n \abs{x}}}{\alpha q_n}\left[B_0 - \sgn(x)B_1\alpha q_n + B_2 (\alpha q_n)^2 + \sgn(x)\tau_z s_z(\alpha q_n)^3\right]\,\tau_z\,s_z\,.
\label{eq:Glowmu}
\end{align}
We can again extract the magnetic field dependence of the proximity-induced gap looking at the energy dependence of the poles in Eq.~\eqref{eq:poles_mu=0}. In general, the minimal gap is dictated by the competition between that of the inner and outer modes. The spectral gap for the inner modes, which we denote $\Delta^{(k=0)}(B)$, is given by
\begin{equation}\label{eq:gap_lowmu}
\Delta^{(k=0)}(B)\begin{cases}
\Delta_0^2\,\sqrt{1-(\tfrac{1}{2}g\mu_B B)^2/\mu^2} & \textrm{if}\;\;\tfrac{1}{2}g\mu_B B<\mu^2/\sqrt{\mu^2+\Delta_0^2}\,,\\
\\
\abs{\sqrt{\Delta_0^2+\mu^2}-\tfrac{1}{2}g\mu_B B} & \textrm{if}\;\;\tfrac{1}{2}g\mu_B B>\mu^2/\sqrt{\mu^2+\Delta_0^2}\,.
\end{cases}
\end{equation}
The gap of the outer modes is not influenced by the magnetic field to the leading order in the ratio $B^2/m\alpha^2$, thus in our effective model it is equal to $\Delta_0$ at all fields. The spectral gap is thus given by
\begin{equation}
\Delta(B) = \min\{\Delta^{(k=0)}(B)\,,\;\Delta_0\}
\end{equation}
At a fixed value of $\mu$, after a slow initial decrease the proximity-induced gap decreases linearly with field and, as already mentioned, closes at $B=B_c(\mu)$, at which point the topological transition takes place (see Fig.~\ref{fig:continuum_gap}).
Increasing $B$ further, the gap $\Delta(B)$ reopens, growing linearly in field until the gap at $k=0$ becomes larger than that at finite momentum.
The gap at finite momentum is equal to $\Delta_0$, while it is well known that this gap has in fact a weak field dependence: it is quadratically suppressed with increasing $B$ if corrections of the order $(g\mu_B B/m\alpha^2)^2$, not included in our approximation, are taken into account.
This limitation is inconsequential for our purposes, since we are mainly interested in the Andreev spectrum in the range of magnetic fields for which the relevant gap is the one at $k=0$.
\end{widetext}

\section{Properties of the Andreev spectrum}
\label{sec:andreev_spectrum}

In this Section we discuss in detail the magnetic field and phase dependence of the Andreev bound state energies.
We begin with a review of the basic notions underpinning the understanding of the excitation spectrum of a Josephson junction.

\subsection{Andreev levels, excitation spectrum, and fermion parity switches}
\label{sec:andreev_spectrum_general_discussion}

Solving the determinant equation derived in the previous Section, Eq.~\eqref{eq:determinant_equation}, allows us to determine the subgap spectrum of the BdG equations \eqref{eq:BdG_equations}.
Since we are dealing with a purely 1D model in the short junction limit, we expect that the subgap spectrum consists of (at most) two distinct Andreev levels.
That is, taking into account the doubling of the spectrum enforced by the particle-hole symmetry, the subgap spectrum of the BdG equations consists of (at most) four solutions $\{\pm E_1, \pm E_2\}$.
Without loss of generality, we fix a hierarchy $0\leq \abs{E_1} \leq \abs{E_2} \leq \Delta(B)$.

\begin{figure}
\begin{center}
\includegraphics[width=0.95\columnwidth]{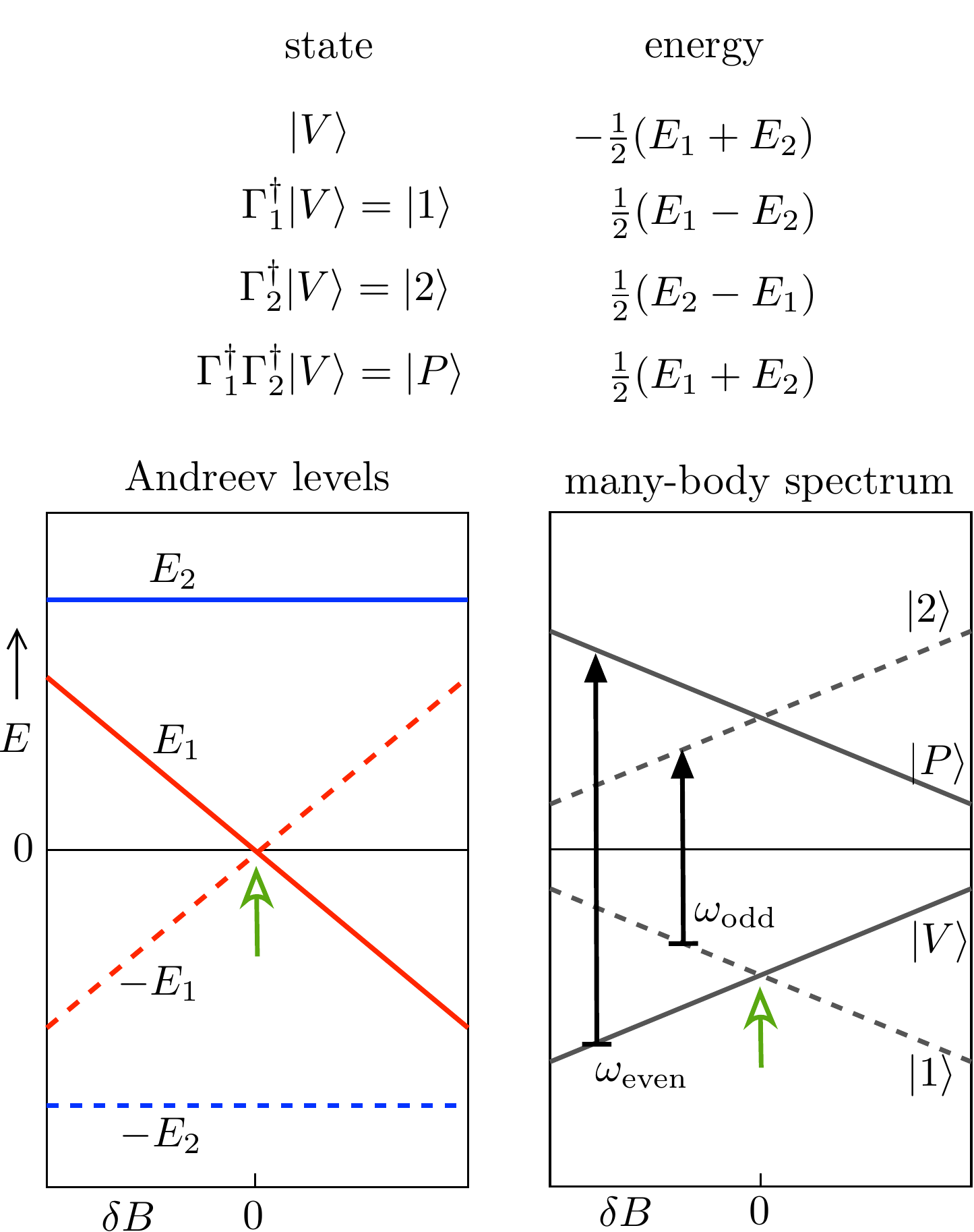}
\caption{Top: energies of the four lowest-lying many-body eigenstates of the junction. The eigenenergies are determined by Eq.~\eqref{eq:H_subgap_expansion}. Bottom: diagrams illustrating the relation between the Andreev levels $\{\pm E_1, \pm E_2\}$ (eigenvalues of the BdG equations, left panel) and the many-body spectrum (right panel). The two diagrams also illustrate the occurrence of a fermion parity switch. We consider a scenario in which, as a parameter of the system is varied, $E_1$ crosses the Fermi level (green arrow in left panel). In the figure, the tuning parameter is represented by the magnetic field close to its switching value, $B=B_\textrm{sw}+\delta B$, see discussion in Sec.~\ref{sec:ABS_in_field}. In the corresponding many-body spectrum, the change of sign of $E_1$ manifests itself as a change in the many-body ground state (green arrow in right panel). The ground state transition is between two states of different fermion parity (assuming the Andreev levels are non-degenerate). In the right panel, the two black arrows mark transition frequencies between states of equal fermion parity, $\omega_\textrm{even}=E_1 + E_2$ and $\omega_\textrm{odd}=E_2-E_1$.}
\label{fig:diagram_spectrum}
\end{center}
\end{figure}

Once the Andreev levels are determined, the many-body Hamiltonian can be expanded as
\begin{equation}\label{eq:H_subgap_expansion}
H = E_1\,(\Gamma^\dagger_1 \Gamma_1 - \tfrac{1}{2}) + E_2\,(\Gamma^\dagger_2\Gamma_2 - \tfrac{1}{2})+ \dots
\end{equation}
where the dots represent the omission of states coming from the continuous part of the spectrum, with energies higher than $\Delta(B)$.
Neglecting the presence of these states, we can limit ourselves to considering just four many-body eigenstates: the vacuum state $\ket{V}$, which is annihilated by both $\Gamma_1$ and $\Gamma_2$; two single-particle states $\ket{1} = \Gamma^\dagger_1\ket{V}$ and $\ket{2} = \Gamma^\dagger_2\ket{V}$; and finally the state with a pair of quasiparticles, $\ket{P}=\Gamma^\dagger_1\Gamma^\dagger_2\ket{V}$.
The fermion parity of the junction, which is a global symmetry of the Hamiltonian, is even in the states $\ket{V}$ and $\ket{P}$, and odd in the states $\ket{1}$ and $\ket{2}$.
Up to a common constant, the energies of these four many-body eigenstates are simply related to the Andreev levels $E_1$ and $E_2$ via Eq.~\eqref{eq:H_subgap_expansion}, see the table in Fig.~\ref{fig:diagram_spectrum}.

Note that, so far, we have not specified the sign of the energies $E_1$ and $E_2$ appearing in Eq.~\eqref{eq:H_subgap_expansion}.
In fact, this choice is arbitrary: as can be seen in Fig.~\ref{fig:diagram_spectrum} the many-body spectrum is invariant under a change of sign of $E_1$ and $E_2$.
This is, again, a consequence of the particle-hole symmetry of the model.
Conventionally, one chooses $E_1$ and $E_2$ to be positive in Eq.~\eqref{eq:H_subgap_expansion}.
In this case, the ground state of the system is identified with the even parity state $\ket{V}$.
The states $\ket{1}$, $\ket{2}$ and $\ket{P}$ are excited states with excitation energies $E_1$, $E_2$ and $E_1 + E_2$ respectively.

Although the initial choice of the sign of $E_1$ and $E_2$ in Eq.~\eqref{eq:H_subgap_expansion} is conventional and does not have measurable consequences, a \textit{change} in the sign of $E_1$ is physical, and it has measurable and important consequences.
Such a change in sign can occur as some of the parameters of the system are varied, typically the magnetic field $B$ or the phase $\phi$.
To fix the ideas, let us assume that $E_1$ is initially positive and that it can be tuned through the point $E_1=0$ by changing a parameter --- a so-called Fermi level crossing (see green arrow on the left panel of Fig.~\ref{fig:diagram_spectrum}).
When $E_1=0$, the states $\ket{V}$ and $\ket{1}$ are degenerate in energy: the energy cost to add a quasiparticle to the junction vanishes (see green arrow on the right panel of Fig.~\ref{fig:diagram_spectrum}).
Furthermore, when $E_1$ becomes negative, the odd-parity state $\ket{1}$ becomes the ground state of the junction.
This ground state transition driven by a Fermi level crossing is commonly referred to as a fermion parity switch.

Fermion parity switches can be generically expected in Josephson junctions with broken time-reversal symmetry \cite{beenakker2013}, and can drastically affect the thermodynamic and transport properties of the junction.
The Yu-Shiba-Rusinov states associated with magnetic impurities in $s$-wave superconductors \cite{yu1965,shiba1968,rusinov1969,balatsky2006} provide an early example of this type of phenomenon.
A fermion parity switch is also at the basis of the $4\pi$-periodic Josephson effect associated with Majoranas \cite{kitaev2001,kwon2004,fu2009}.
In this case, the peculiarity is that there is an odd number of fermion-parity switches in a $2\pi$ phase interval, a signature of the presence of a fermion-parity anomaly in the low-energy theory of the junction (only an even number of fermion parity switches in a $2\pi$ phase interval is allowed in a topologically trivial phase).
Later in this Section, we will investigate the occurrence of fermion parity switches in the model under study, both in the trivial and topological phases.
Before doing so, we provide an overview of the features of the Andreev spectrum of the model, starting from the well-known case in which $B=0$.

\subsection{Solution at zero magnetic field}
\label{sec:SolutionAtZeroField}

At zero magnetic field, an analytic solution leads to a well-known universal result for the Andreev levels \cite{beenakker1991,furusaki1991,furusaki1999}.
The Andreev levels form a degenerate doublet, $E_1=E_2\equiv E_A$ with
\begin{equation}\label{eq:EA}
E_A = \Delta_0\,\left[1-\tau\sin^2(\phi/2)\right]^{1/2}\,.
\end{equation}
This result is valid independently on the values of chemical potential $\mu$ and spin-orbit coupling $\alpha$, provided that the Andreev approximation is applicable.

While the solution \eqref{eq:EA} is already well-known, it is instructive to reproduce this result from Eq.~\eqref{eq:determinant_equation}.
At $B=0$, the Green's function $G(0^-,E)$, which can be deduced from Eqs.~\eqref{eq:Ghighmu} or \eqref{eq:Glowmu}, takes a particularly simple form:
\begin{equation}\label{eq:G0_at_zero_field}
G(0^-, E)=\frac{i}{2}\frac{\Delta_0}{\sqrt{\Delta_0^2-E^2}}\,\left[\tau_x - \e^{i \beta(E) \tau_z s_z}\right]\,\tau_z s_z\,,
\end{equation}
with $\beta(E)=\arccos(E/\Delta)$. There are three meaningful facts about the above expression.
First, it is valid for both limits $\mu\gg m\alpha^2, \Delta_0$ and $m\alpha^2\gg \mu, \Delta_0$, so we already see that the solutions of the determinant equation~\eqref{eq:determinant_equation} will be common to the two cases.
Second, in both limits the right hand side of Eq.~\eqref{eq:G0_at_zero_field} is independent of the spin-orbit coupling strength $\alpha$.
This is a consequence of the fact that spin-orbit coupling can be removed from the Hamiltonian via a local gauge transformation, and so the Green's function evaluated at a single point can be made independent of $\alpha$.
Thus, the independence of $E_A$ on $\alpha$ can also be explained as a consequence of the short junction limit.
Third, the right hand side of Eq.~\eqref{eq:G0_at_zero_field} is proportional to the unit matrix in the spin grading, which leads to the anticipated double degeneracy of the solutions.
Plugging the Green's function from Eq.~\eqref{eq:G0_at_zero_field} and the transfer matrix from Eq.~\eqref{eq:transfer_matrix_with_tau} into the determinant equation \eqref{eq:determinant_equation}, we obtain the solution \eqref{eq:EA}.

It is also possible to write down the bound state wave functions explicitly.
In order to do this, the first step is to solve the system of linear equations $G(0^-, E_A)\,M\,\tilde\Phi(0^-)=\tilde\Phi(0^-)$ to find the wave functions at the position $x=0^-$.
Then, using the knowledge of $G(x, E_A)$ at arbitrary $x$, one can reconstruct the entire wave function using Eq.~\eqref{eq:wavefunction}.
Carrying out this procedure, one finds two solutions $\Phi_1(x)$ and $\Phi_2(x)$, which are written out in detail in Appendix~\ref{app:zero_field_bound_state_wavefunctions}.
In our model, spin along the $z$ direction is a good quantum number at $B=0$, and $\Phi_1(x)$ and $\Phi_2(x)$ are identical except for the fact that they carry opposite spin.
As anticipated in the previous paragraph, the spin-orbit interaction is not effective in separating the two Andreev levels with opposite spins in energy.
This would be true even in a model where the spin-orbit interaction takes a more general form and breaks the spin rotation symmetry completely.
The degeneracy can not be explained by invoking Kramers' theorem either: the Kramers partner of $\Phi_1(x,\phi)$ is $\Phi_2(x,-\phi)$, so that the two wave functions form a true Kramers doublet only at the time-reversal invariant points $\phi=0,\pi$.
Rather, the degeneracy of the Andreev levels is a consequence of the short-junction limit.
It is removed by spin-orbit coupling if corrections of order $(L/\xi)$ are taken into account \cite{chtchelkatchev2003,beri2008}, or even in the short junction limit in the case of a multi-terminal junction \cite{heck2014}.

\subsection{Magnetic field dependence of the spectrum: qualitative features}
\label{sec:ABS_in_field}

When a finite magnetic field is present, in general we find that the Andreev level spectrum cannot be found analytically.
Thus, away from simple limits, we resort to a numerical search of the roots of the determinant Eq.~\eqref{eq:determinant_equation}.
In total, once one of the two linearization limits is taken, there are four parameters which determine the spectrum: the magnetic field $B$, the phase $\phi$, the transparency of the junction $\tau$, and either the spin-orbit coupling $\alpha$ (when $\mu \gg \Delta_0, m\alpha^2, \tfrac{1}{2}g\mu_B B$) or the chemical potential $\mu$ (when $m\alpha^2\gg \Delta_0, \mu, \tfrac{1}{2}g\mu_B B$).
We focus in particular on the field and phase dependence of $E_1$ and $E_2$, since these are the two parameters which are varied systematically in experiment.

\begin{figure*}[t!]
\begin{center}
\includegraphics[width=\textwidth]{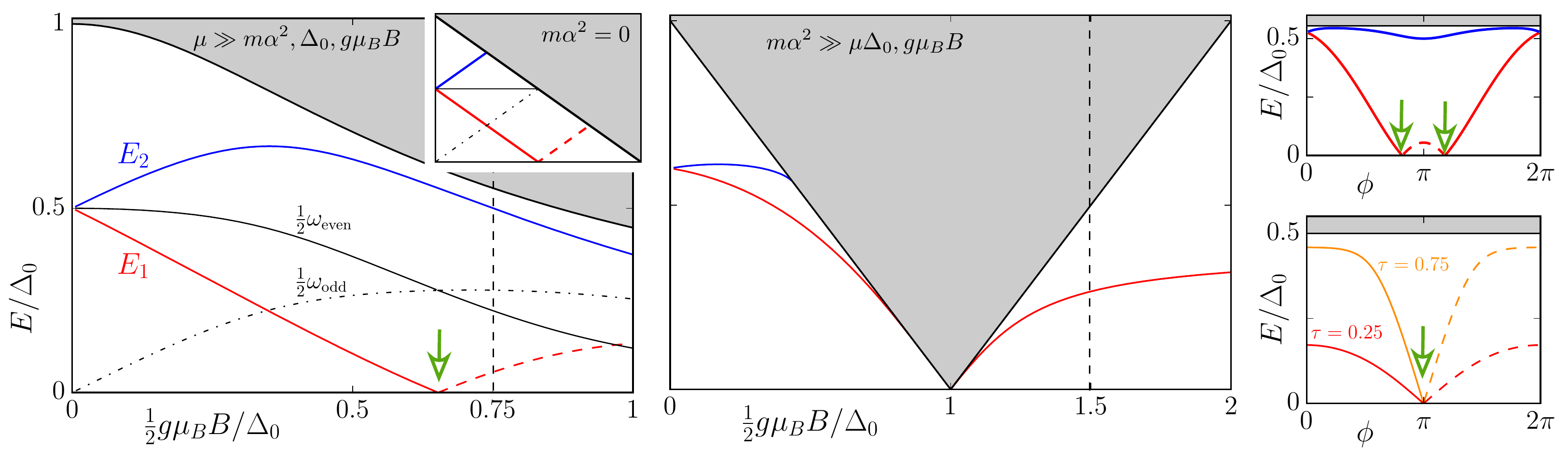}
\caption{\emph{Left panel:} magnetic field dependence of the Andreev levels $E_1$ (red) and $E_2$ (blue) for $\mu\gg \Delta_0, m\alpha^2, g\mu_B B$ at a fixed value of the phase $\phi=\pi$. Other parameters: $\tau=0.75$, $\alpha k_F/\Delta=0.5$. The solid black line is the gap $\Delta(B)$ of the continuum (gray area). The thin black lines represent the transition frequencies $\omega_\textrm{even}$ and $\omega_\textrm{odd}$ which determine the microwave absorption spectrum (the factor $\tfrac{1}{2}$ is included for convenience). Note the presence of a fermion parity switch (green arrow) at which $E_1=0$ and $\omega_\textrm{even}=\omega_\textrm{odd}$. For comparison, the inset shows the equivalent spectrum obtained in the absence of spin-orbit coupling, $\alpha=0$. \emph{Middle panel:} magnetic field dependence on the Andreev levels in the low chemical potential regime, with parameters $\mu=0$, $\phi=3\pi/4$ and $\tau=0.75$. As in the left panel, the solid black line is the gap $\Delta$ of the continuum, which vanishes at the topological transition. The single Andreev level appearing in the topological phase after the gap closing originates from the hybridization of two Majorana zero modes at the junction. \emph{Right panels:} Phase dependence of the Andreev levels at the value of the $B$ marked by the vertical dashed lines in the left and middle panels, for both trivial and topological phases (upper and lower panels respectively). The bottom panel shows the phase dependence of the Andreev level in the topological phase for two different values of the transmission probability $\tau$. The green arrow marks the point of a single Fermi level crossing at $\phi=\pi$, which is at the origin of the $4\pi$-periodic Josephson effect marking the topological nature of the high-field phase. By contrast, in the trivial phase Fermi level crossing always come in pairs in a $2\pi$ interval, see the upper panel.\label{fig:abs_spectra}}
\end{center}
\end{figure*}

Let us first discuss the simple situation in which spin-orbit coupling is absent, $\alpha=0$.
In this case, spin is a good quantum number and the Zeeman interaction is separable from the rest of the Hamiltonian.
One simply obtains a linear Zeeman splitting, $E_1 = E_A - \tfrac{1}{2}g\mu_B B$ and $E_2 = E_A + \tfrac{1}{2}g\mu_B B$, with the same g-factor as that of the continuum states (see inset in the right panel of Fig.~\ref{fig:abs_spectra}).
Note that by increasing the magnetic field one reaches a field value $B_\textrm{sw}(\phi)= 2 E_A(\phi)/g\mu_B$ at which $E_1$ changes sign: a Fermi level crossing occurs.
Because $E_A(\phi)$ has a minimum at $\phi=\pi$, this is the value of the phase at which the Fermi level crossing occurs first upon increasing the magnetic field.
After this point, i.e. for $B>B_\textrm{sw}(\pi)$, a pair of fermion parity switches is nucleated symmetrically around $\phi=\pi$ (see for instance the top right panel of Fig.~\ref{fig:abs_spectra}).
This behavior is consistent with the fact that, in a topologically trivial phase, the number of fermion parity switches in a $2\pi$ phase interval must be even.
While $E_1$ decreases with field, the other Andreev level $E_2$ increases and merges with the continuum of states with opposite spins at a field value $B_\textrm{cross}=2(\Delta_0-E_A)/g\mu_B$.
This crossing of the Andreev level with the continuum is protected by spin conservation.

The magnetic field dependence of Andreev level spectrum is qualitatively different in the presence of spin-orbit coupling.
The typical behavior of the Andreev spectra at fixed $\tau, \phi$ and $\alpha$ is shown in Fig.~\ref{fig:abs_spectra} in both linearization limits.
Before entering into the quantitative details of the features of the Andreev levels, let us discuss the important qualitative features.

We begin by discussing the case $\mu\gg m\alpha^2, \Delta_0, g\mu_BB$, illustrated in the left panel of Fig.~\ref{fig:abs_spectra}.
For small magnetic fields, the two Andreev levels $E_1$ and $E_2$ split linearly.
The lowest-lying level $E_1$ maintains its approximately linear behavior in $B$ up to the occurrence of a Fermi level crossing.
Similarly to the zero spin-orbit coupling case discussed earlier, Fermi level crossings first appear at $\phi=\pi$ upon increasing the magnetic field and are then nucleated in pairs around this point.
The field $B_\textrm{sw}(\pi)$ at which the Fermi level crossing first occurs depends on $\alpha$ and $\tau$: this dependence is investigated in detail later.
The energy $E_2$ of the second Andreev level increases with $B$, but bends down at $B\gtrsim B_\textrm{cross}$, when $E_2$ becomes close in energy to the continuum gap $\Delta(B)$, which is decreasing in field.
This is due to the fact that, in the presence of both Zeeman and spin-orbit couplings, there are no symmetries in the model which protect the crossing of the Andreev level with the continuum.
This avoided crossing between the Andreev level and the continuum leads to a non-monotonic dependence of $E_2$ on $B$.
Such a non-monotonic dependence is the cause of the suppression in $B$ of the transition frequency $\omega_\textrm{even}=E_1+E_2$ between the two junction states with even parity, a fact which we used to explain the observed absorption spectra of an InAs/Al Josephson junction in Ref.~\cite{vanwoerkom2016} (see also Sec.~\ref{sec:absorption}).

In the low chemical potential regime, shown in the middle panel of Fig.~\ref{fig:abs_spectra}, the two Andreev levels also split linearly for small magnetic fields.
However, their behavior at large fields is drastically different from that at high chemical potential, due to the different behavior of the gap $\Delta(B)$.
The two Andreev levels merge in rapid sequence with the continuum of states -- whose gap is linearly decreasing -- right before the topological transition at $B=B_c$.
In the topological phase at $B>B_c$, we find that the subgap spectrum consists of a single pair of Andreev levels $\pm E_1$.
We may see the energy $E_1$ as the result of the coupling between two Majorana zero modes located at the two interfaces of the junction.
This notion is accurate in particular for $\tau\ll 1$, when the two interfaces are weakly coupled.

The phase dependence of $E_1$ in the topological phase is shown in the bottom right panel of Fig.~\ref{fig:abs_spectra} for two different values of the junction transparency $\tau$.
In both cases, and for any $B>B_c$, the energy spectrum displays a single Fermi level-crossing at $\phi=\pi$.
(This behavior should be contrasted with that of the topologically trivial phase, where, as discussed earlier, Fermi level crossings  appear in pairs, see top right panel of Fig.~\ref{fig:abs_spectra}).
The pinning of the position of the Fermi level crossing at $\phi=\pi$ for $B>B_c$ is due to a symmetry of our particular model.
Under the combined operation $\mathcal{S}=\sigma_x \mathcal{R}$, where $\mathcal{R}$ is the operator of spatial inversion $x\mapsto -x$, the Hamiltonian in Eq.~\eqref{eq:wire_model} is mapped to itself up to the change $\phi\mapsto -\phi$ \cite{mengcheng2012}.
This dictates that the Andreev spectrum must be symmetric around $\phi=0$, i.e. that $E_1(\phi)=E_1(-\phi)$.
If, additionally, we recall that the entire spectrum must be $2\pi$ periodic in $\phi$, the only allowed point where $E_1$ can vanish is indeed $\phi=\pi$ (this consideration holds in the case that only a single Fermi level crossing is present in a $2\pi$ period.)
A Josephson junction with more transport channels or a denser Andreev spectrum may exhibit a higher number of Fermi level crossings \cite{law2011}, and in a model where there are no constraints coming from spatial inversion the position of the Fermi level crossing may be in general different from $\pi$.

In the rest of this Section, we investigate in more detail the different qualitative features of the Andreev level spectrum described so far.

\subsection{Behavior at small field: Zeeman splitting of the Andreev levels}
\label{sec:g-factor}

We have seen that in both linearization limits the Andreev levels split starting from infinitesimally small magnetic fields.
The linear-in-$B$ splitting can be captured by standard degenerate perturbation theory applied to the zero-field wave functions presented in Appendix~\ref{app:zero_field_bound_state_wavefunctions}.
This procedure is valid as long as $|\tfrac{1}{2}g\mu_B B|\ll \Delta_0 - E_A$, so that the discrete Andreev levels are distant from the continuum part of the spectrum. 
Thus, the results presented in this section are most relevant for $1-\tau\ll1$ and $\abs{\phi-\pi}\ll \pi$, i.e. when the energy $E_A$ is much lower than the gap $\Delta_0$.

It is useful to cast the result of the perturbation calculation in terms of an effective g-factor which is the linear coefficient of the expansion of $E_1$ and $E_2$ around $B=0$,
\begin{align}
E_1& = E_A - \tfrac{1}{2}\,g_A\,\mu_B\,B + \dots\\
E_2& = E_A + \tfrac{1}{2}\,g_A\,\mu_B\,B + \dots
\end{align}
We find that the Andreev level g-factor $g_A$ is different from the ``bare'' value, $g$, which determines the size of the Zeeman gap at $k=0$ in the homogeneous wire, and that $g_A$ can depend strongly on the system parameters.

At perfect transmission, $\tau=1$, the zero-field Andreev bound state wave functions are eigenstates of the velocity operator $s_z$ [see Eq.~\eqref{eq:wavefzerofield}].
Therefore, only the part of the Zeeman coupling which mixes co-propagating modes is effective in splitting the Andreev levels (see Appendix~\ref{app:derivation_gfactor} for a discussion).
In this case, it is possible to derive an expression for $g_A$ which is valid at any value of the ratio $\mu/m \alpha^2$, provided that $\max(\mu, m\alpha{}^2)\gg \Delta_0$:
\begin{equation}\label{eq:g_factor_perfect_transmission}
\frac{g_A}{g} = \frac{\Delta_0^2\,\sin^2(\phi/2)}{\Delta_0^2 \sin^2(\phi/2) + m\alpha^2\,(2\mu+m\alpha^2)}\,,\;\;(\tau=1).
\end{equation}
The equation above can be derived by using a linearization procedure which interpolates between the two  limits $\mu\gg m\alpha^2$ and $m\alpha^2\gg \mu$ used in Sec.~\ref{subsec:linearization_mu} and \ref{subsec:linearization_mu=0} respectively; the derivation is contained in Appendix~\ref{app:derivation_gfactor}.
When $m\alpha^2=0$, Eq.~\eqref{eq:g_factor_perfect_transmission} yields $g_A=g$ independently on the value of all other parameters.
At any finite value of $m\alpha^2$ the Andreev bound state g-factor $g_A$ is reduced with respect to the bare value, $g$.
The suppression is the strongest when $m\alpha^2\gg \mu, \Delta_0$, in which case Eq.~\eqref{eq:g_factor_perfect_transmission} yields $g_A\ll g$.
A finite spin-orbit coupling also makes $g_A$ dependent on the phase difference $\phi$, with a maximum at $\phi=\pi$.

The considerations in the previous paragraph, based on Eq.~\eqref{eq:g_factor_perfect_transmission}, remain qualitatively valid also for $\tau<1$.
In the presence of scattering, the Andreev bound state wave functions are superpositions of states with opposite velocity.
In this case, the magnetic field mixes the counter-propagating components (originating from modes close to $k=0$) as well as the co-propagating ones (originating from modes at finite $k$) - see the discussion in Appendix~\ref{app:derivation_gfactor}.
We may write the Andreev level g-factor as a sum of two terms, $g_A=g_\rightrightarrows+g_\rightleftarrows$, corresponding to these two different contributions.
The co-propagating contribution is given by
\begin{equation}\label{eq:g_copropagating}
\frac{g_\rightrightarrows}{g} = \frac{\Delta_0^2-E_A^2}{\Delta_0^2-E_A^2 + m\alpha^2\,(2\mu+m\alpha^2)}\,,
\end{equation}
of which Eq.~\eqref{eq:g_factor_perfect_transmission} is a special case.
Equation~\eqref{eq:g_copropagating} is the dominant contribution to the g-factor when $\mu\gg m\alpha^2, \Delta_0$, in which case $\abs{g_\rightrightarrows}\gg \abs{g_\rightleftarrows}$ and $g_A\approx g_\rightrightarrows$ for any value of the transmission $\tau$.

\begin{figure}[t!]
\begin{center}
\includegraphics[width=\columnwidth]{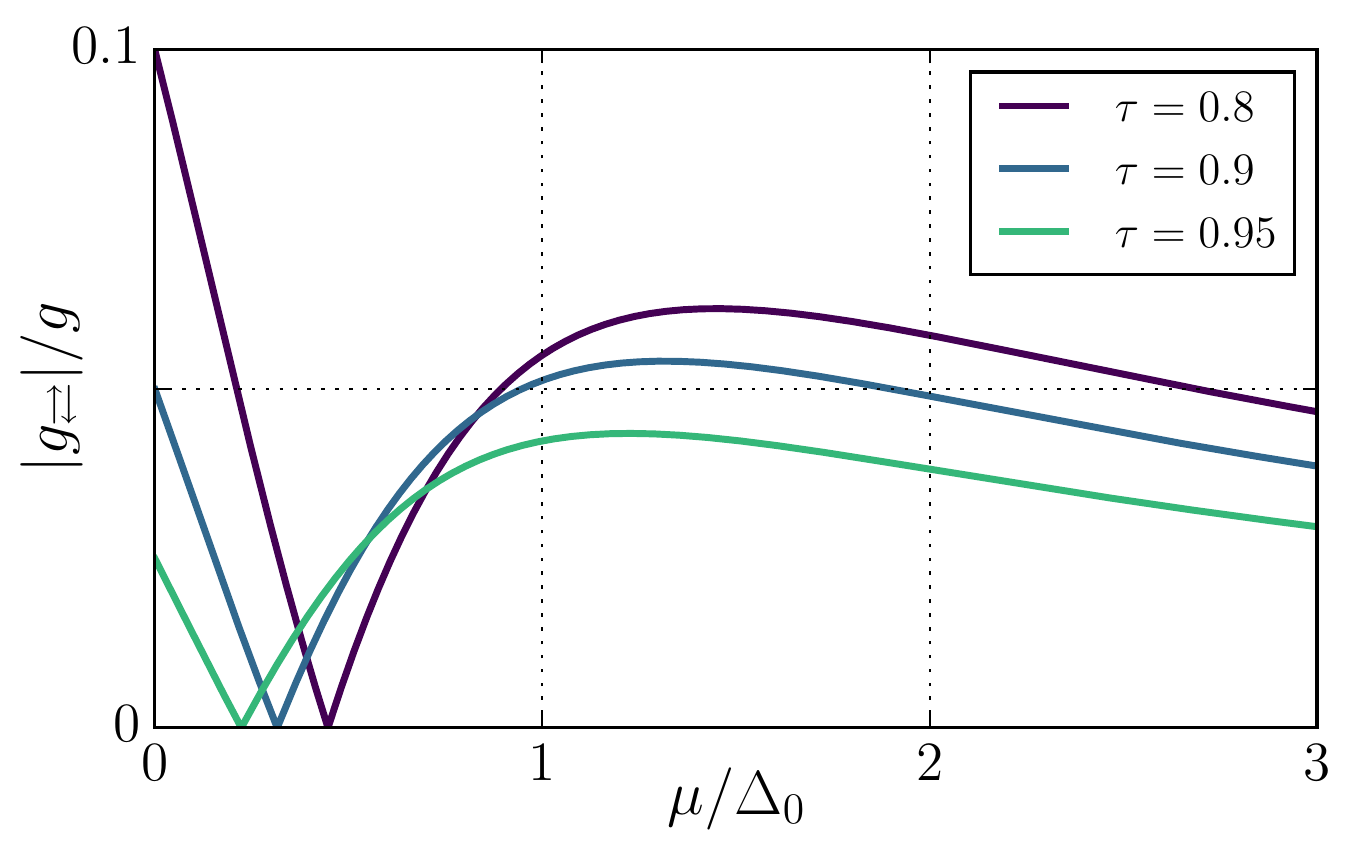}
\caption{Contribution $g_\leftrightarrows$ to the Andreev level g-factor $g_A$ coming from the coupling between counterpropagating modes, normalized to the bare g-factor $g$ of Eq.~\eqref{eq:wire_model}. The curves are obtained from Eq.~\eqref{eq:g_counterpropagating} with $\phi=\pi$ for different values of $\tau$. The zeros of $g_\rightleftarrows$ happen at $\mu_0=\Delta_0\sqrt{1-\tau}\abs{\sin(\phi/2)}$. Away from these points, $g_\leftrightarrows$ provides the leading contribution to the Andreev g-factor $g_A$ at $\mu,\Delta_0\ll m\alpha^2$. Note that, as explained in the main text, $g_A$ may be much smaller than the bare g-factor $g$ of Eq.~\eqref{eq:wire_model}.}
\label{fig:g_factor_counterpropagating}
\end{center}
\end{figure}

The counter-propagating contribution $g_\rightleftarrows$ becomes relevant in the opposite regime $m\alpha^2 \gg \mu, \Delta_0$.
Indeed, in this regime the dominant mixing introduced by a small magnetic field is the one between the counter-propagating modes at $k=0$, which both participate in the formation of the Andreev bound states provided that $\tau<1$.
In the limit $\mu/m\alpha^2\to 0$ and $\Delta_0/m\alpha^2\to 0$, the one treated in Sec.~\ref{subsec:linearization_mu=0}, we find
\begin{align}\label{eq:g_counterpropagating}
\frac{g_\rightleftarrows}{g} &= \frac{\tau\sqrt{1-\tau}}{2}\frac{\mu \Delta_0 \abs{\sin(\phi/2)}}{\Delta_0^2-E_A^2+\mu^2}-\frac{1-\tau}{2}\frac{\Delta_0^2-E_A^2}{\Delta_0^2-E_A^2+\mu^2}\,,
\end{align}
which is illustrated in Fig.~\ref{fig:g_factor_counterpropagating}.
Equation~\eqref{eq:g_counterpropagating} is the leading contribution to the total g-factor $g_A = g_\rightrightarrows+g_\rightleftarrows$ at low chemical potential, except for the vicinities of $\tau= 1$ and $\mu=\mu_0$, with $\mu_0=\Delta_0\sqrt{1-\tau}\abs{\sin\phi/2}$.
In these narrow regions of the parameter space, Eq.~\eqref{eq:g_counterpropagating} is vanishing and thus the g-factor is determined by the co-propagating contribution $g_\rightrightarrows$, in spite of its smallness.
Furthermore, note that $g_\rightrightarrows$ and $g_\rightleftarrows$ have competing signs when $0<\mu<\mu_0$, and so in this region higher-order corrections in the parameter $\mu/m\alpha^2$ may be crucial to determine the g-factor (including its overall sign).
However, as discussed in Appendix~\ref{app:derivation_gfactor}, the correction to Eq.~\eqref{eq:g_counterpropagating} due to a finite ratio $\mu/m\alpha^2$ cannot be easily computed within a linearized spectrum approximation, since such a calculation necessarily involves the electronic state close to the bottom of the parabolic bands of Fig.~\ref{fig:linearization_limits}.
Finally, we note that Eqs.~\eqref{eq:g_copropagating} and \eqref{eq:g_counterpropagating} agree in predicting a $\sim1/\mu$ suppression of $g_A$ when $\mu\gg \Delta_0$.

The value of $g_A$ is not directly accessible in microwave absorption spectroscopy, since the microwave transition frequency $\omega_\textrm{even}=E_1+E_2$ is insensitive to the linear splitting in $B$.
However, it is observable in tunneling spectroscopy, which can access $E_1$ and $E_2$ individually.
The analysis contained in the above paragraphs suggests that a systematic investigation of $g_A$ may be valuable to obtain information about the electron density and the strength of the spin-orbit coupling in the nanowire.
This investigation can be carried out at very small values of the field and may be helpful in predicting or understanding the high-field behavior of the system.

\subsection{Occurrence and position of Fermi level crossings}
\label{sec:fermi_level_crossings}

\begin{figure}[t!]
\begin{center}
\includegraphics[width=\columnwidth]{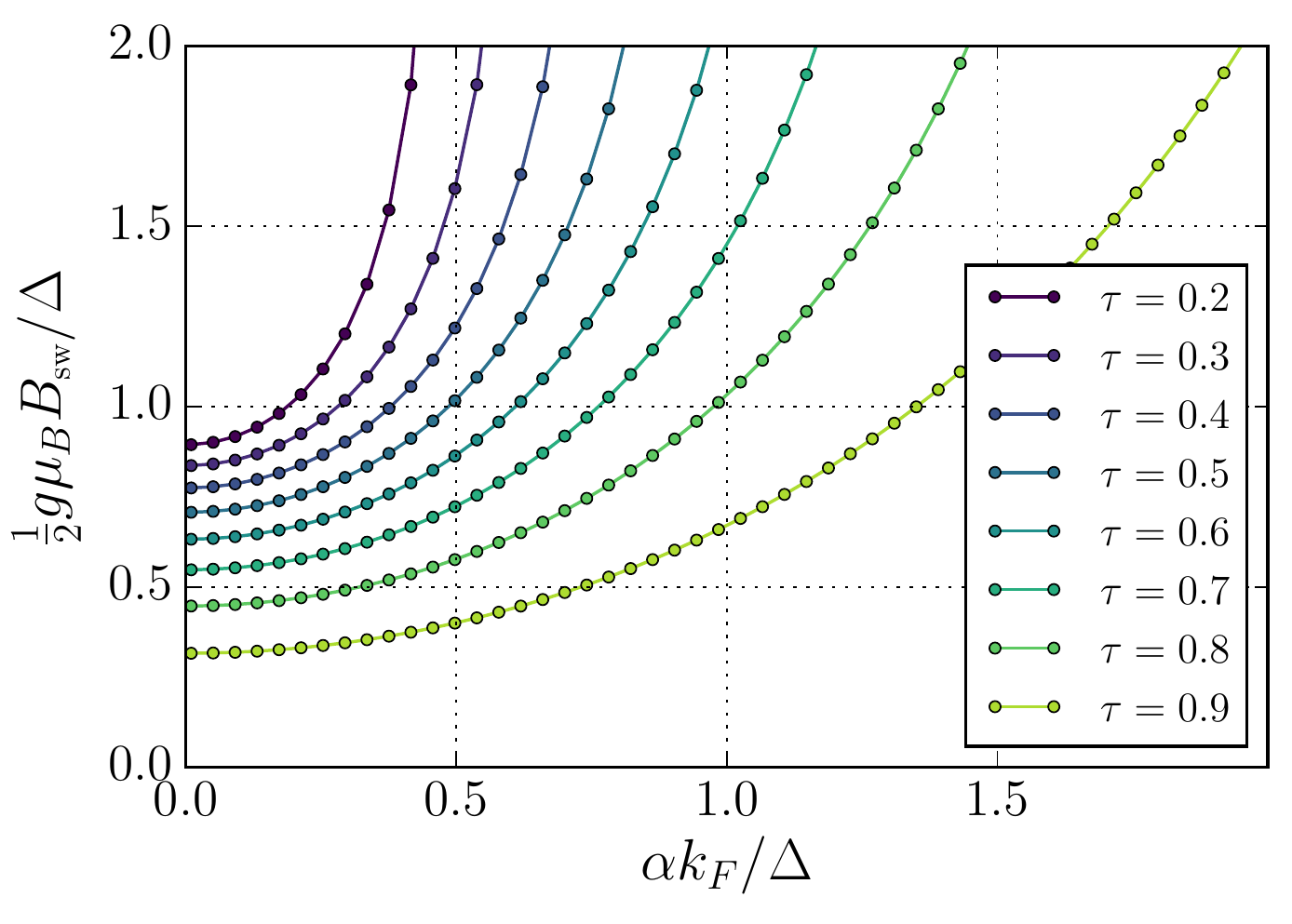}
\caption{Dependence of the magnetic field $B_\textrm{sw}$ -- at which a Fermi level crossing occurs -- on the spin-orbit strength, for different values of the transmission coefficient $\tau$ and at fixed phase $\phi=\pi$, for the high chemical potential regime $\mu\gg \Delta_0$.\label{fig:b_sw_vs_spin-orbit}}
\end{center}
\end{figure}

Earlier in the text, we have seen that, in the high chemical potential regime $\mu\gg m\alpha^2, g\mu_B B, \Delta_0$, Fermi level crossings may occur at a field $B=B_\textrm{sw}$ (see the left panel of Fig.~\ref{fig:abs_spectra}).
In Fig.~\ref{fig:b_sw_vs_spin-orbit} we study in more detail the dependence of $B_\textrm{sw}$, computed at $\phi=\pi$, on spin-orbit coupling strength and transmission.

There are two notable trends.
First, the switching field $B_\textrm{sw}$ decreases upon increasing the transmission $\tau$ at fixed spin-orbit strength.
This is due to the fact that the larger $\tau$ is, the closer to zero is $E_A$, and thus a smaller field is required to induce the Fermi level crossing.
Second, when increasing the spin-orbit strength at fixed transmission, the field $B_\textrm{sw}$ increases.
This is due to the suppression of the Andreev level g-factor $g_A$ with increasing spin-orbit strength or chemical potential, see Eq.~\eqref{eq:g_copropagating}, which leads to a slower decrease of $E_1$ with $B$.
Our numerical results suggest that there is a value $(\alpha k_F)_\textrm{max}$ above which the Fermi level crossings are absent: that is, the curves in Fig.~\ref{fig:b_sw_vs_spin-orbit} have an asymptote at finite $\alpha k_F$ at which $B_\textrm{sw}$ diverges.
Qualitatively, a strong spin-orbit coupling may prevent the Fermi level crossing to occur because of the level repulsion between the Andreev level $E_1$ and the negative image of the rest of the spectrum.
Judging from the numerical data shown in Fig.~\ref{fig:b_sw_vs_spin-orbit}, $(\alpha k_F)_\textrm{max}$ depends on the transmission $\tau$, and it grows with increasing $\tau\to 1$.
We attribute this behavior to the fact that, in the limit $\tau\to 1$, $E_A(\pi)\to0$: thus, a Fermi level crossing appears already at an infinitesimally small field, and it becomes prohibitive to remove it.
Finally, we notice that numerical calculations do not reveal the presence of Fermi level crossings in the opposite regime $m\alpha^2\gg \mu, g\mu_B B, \Delta_0$.
We attribute this behavior to the fact that, in this regime, $g_A\ll g$.
Therefore, the decrease in energy of the Andreev bound states is much slower than that of the continuum states (see for instance the middle panel of Fig.~\ref{fig:abs_spectra}), preventing the occurrence of a Fermi level crossing at a field $B<B_c$.

In a tunneling spectroscopy experiment, the closing of the excitation gap of the junction at a Fermi level crossing in the regime $\mu\gg \Delta_0$ may be naively mistaken for a bulk topological transition.
Indeed, a typical magnetic field scale for a fermion parity switch is $B_\textrm{sw}\sim500$ mT \cite{vanwoerkom2016}, not dissimilar from that of the critical field $B_c$ \cite{albrecht2015}.
The strong dependence of $B_\textrm{sw}$ on $\tau$ (as well as $\phi$), however, should allow to discriminate easily between the two cases.

\section{Current operator and the equilibrium current}
\label{sec:current}

In this Section, we evaluate the temperature and magnetic field dependence of the equilibrium current.
It is known that, in a short Josephson junction not subject to magnetic field, the current is carried almost entirely by the Andreev bound states \cite{beenakker1991,furusaki1999,heikkila2002,levchenko2006}. 
This conclusion remains true also in the presence of magnetic field (with or without spin-orbit coupling), as we will now argue following the discussion from Ref.~\cite{levchenko2006}. 
On one hand, the energies of the Andreev bound states vary by an amount $\sim\Delta$ upon varying the phase $\phi$, and thus they provide a finite contribution to the current in the limit $L/\xi\to 0$.
On the other hand, the contribution of the continuous spectrum to the current density comes from states within the energy range $\Delta < E < E_\textrm{Th}$.
Here, $E_\textrm{Th}$ is the Thouless energy, i.e. the energy scale associated with the flight time of quasiparticles across the junction; in a short quasi-ballistic junction, the Thouless energy is large, $E_\textrm{Th}/\Delta\sim\xi/L\gg 1$.
The spectral density of the current delivered by states with energy $\sim E$ scales as $\Delta^2/(E_\textrm{Th}E)$ for energies in the interval $E_\textrm{Th}\gtrsim E\gg\Delta$.
It yields a total contribution $\propto (\L/\xi)\,\ln(\xi/L)$ to the current, which vanishes in the limit $L/\xi \to 0$ \cite{levchenko2006}. 
This argument remains valid even in the presence of a magnetic field or spin-orbit coupling.
Therefore, in the following we neglect the contribution of the extended states to the current. 

We start by finding the current operator $j(x)$ for the junction, and then evaluate the contribution of the many-body eigenstates $|V \rangle,\, |1 \rangle,\, |2 \rangle, $ and $|P \rangle$, see Fig.~\ref{fig:diagram_spectrum}. 
The current operator can be derived from a continuity equation for the electric charge density $\rho$, which for the original model of Eq.~\eqref{eq:wire_model} is given by the operator  
\begin{equation}
\rho(x) =  \frac{e}{2}\,\psi^\dagger(x) \tau_z \psi(x)\,.
\end{equation}
The continuity equation for $\rho$ can be computed using the equation of motion of the field $\psi(x)$ under the Hamiltonian of Eq.~\eqref{eq:wire_model}.
It can be cast in the form
\begin{equation}
\pd_t\,\rho(x) + \pd_x\,j(x) = s(x)\,,
\end{equation}
with $j(x)$ the quasiparticle current operator, which includes a contribution from the spin-orbit coupling,
\begin{equation}
j(x) = \frac{e}{2mi}  \psi^\dagger(x)\,\partial_x \psi(x)  + \frac{e}{2}\,\alpha\,\psi^\dagger(x)\,\sigma_z\,\psi(x)\,,
\end{equation}
and $s(x)$ a charge source (or drain) term due to the presence of the superconducting condensate \cite{blonder1982},
\begin{equation}
s(x) = e \Delta_{0}\,\psi^{\dagger}(x)\,\tau_{y}e^{-i\phi\,\mathrm{sgn}(x)\tau_{z}/2}\,\psi(x)\,.
\end{equation}
At the position of the junction, $x=0$, the source term vanishes since there is no proximity-induced pairing $\Delta_0$.
Thus, at the junction the equilibrium current can be computed by studying only the quasiparticle contribution coming from $j(x)$.
In the superconducting leads, the quasiparticle current is converted into current carried by the condensate over a length $\sim\xi$. Correspondingly, the contribution of the $j(x)$ term to the equilibrium current decays away from $x=0$. The decay is compensated by the source term \cite{blonder1982} to ensure the current conservation along the wire.

Using Eq.~(\ref{eq:linearized_field_kF}) and Eq.~(\ref{eq:linearized_field_mu=0}) we find the current operator projected to low energies [recall that $\Psi = (\psi_R\,,\,\psi_L)^T$ encodes the left- and right-moving envelope fields],
\begin{equation}\label{eq:current_proj_kF}
j(x) = \frac{ev}{2} \Psi^\dagger(x) s_z \Psi(x)\,,
\end{equation}
where $v = v_F $ in the limit of high chemical potential, and $v = \alpha$ for $m\alpha^2\gg \mu, g\mu_BB, \Delta_0$.
Using the pseudounitarity of the transfer matrix, Eq.~(\ref{eq:pseudounitary_transfer_matrix}), together with the boundary condition for $\Psi$ at the origin, Eq.~\eqref{eq:boundary_condition_Psi}, one can check that the linearized current operator in Eq.~\eqref{eq:current_proj_kF} is continuous across the junction, i.e. $j(0^-) = j(0^+)$. We will thus evaluate the current at $x=0^-$ from now on and omit the position argument.

\begin{figure}[t!]
\begin{center}
\includegraphics[width=\columnwidth]{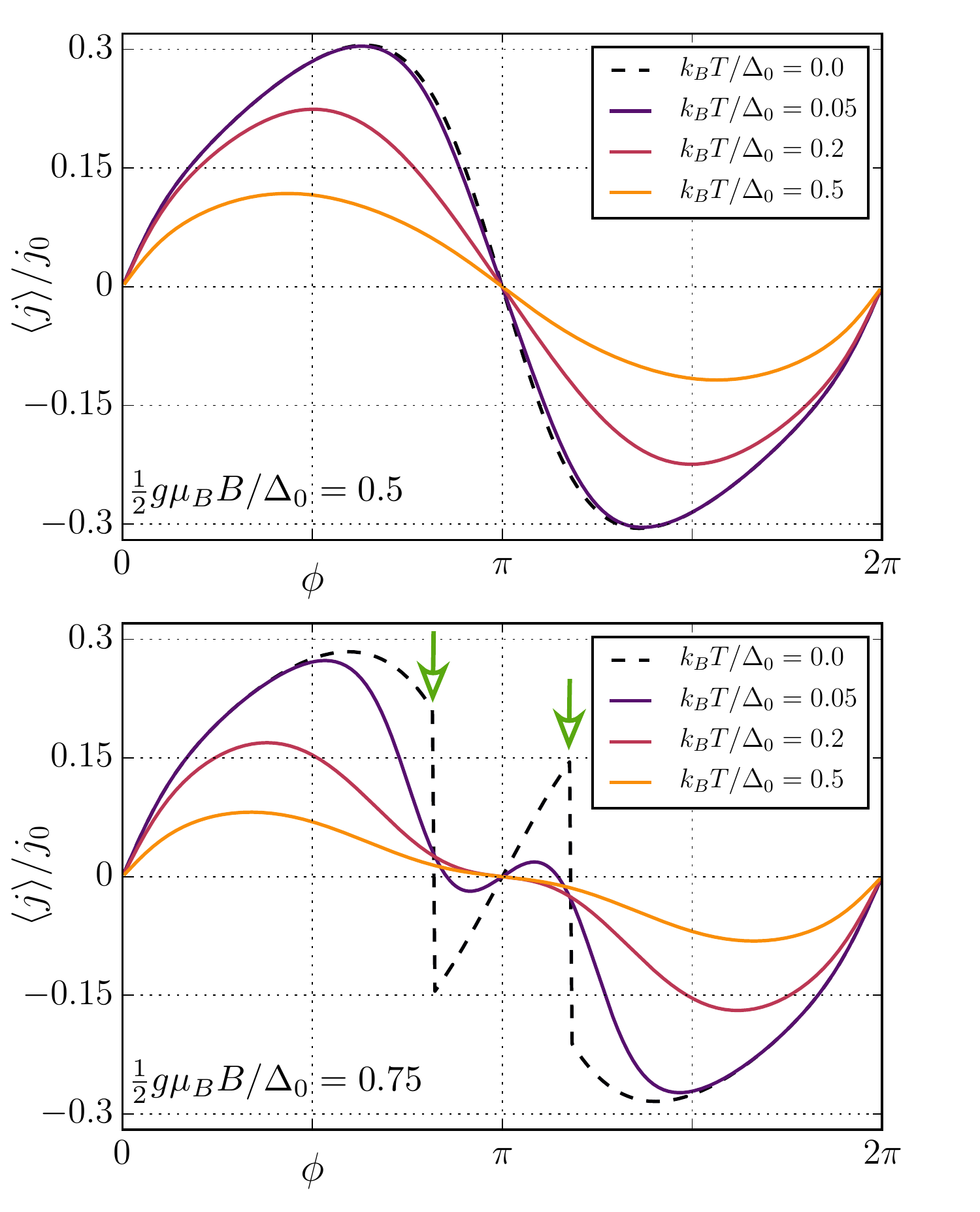}
\caption{Current-phase relation of the nanowire Josephson junction at equilibrium at different temperatures. The current is measured in units of $j_0=e\Delta_0/\hbar$. The equilibrium current is computed in the regime $\mu\gg \Delta_0, g\mu_B B, m\alpha^2$ for two different values of magnetic field $B$ (upper and lower panels) and with all other parameters as in the left panel Fig.~\ref{fig:abs_spectra}: $\alpha k_F/\Delta_0=0.5, \tau=0.75$. The two values of magnetic field are chosen to be on the left and on the right of the fermion parity switch in the left panel of Fig.~\ref{fig:abs_spectra}. The bottom panel thus reveals the effect of fermion parity switches (marked again by green arrows) on the equilibrium current. At $T=0$, fermion parity switches are signaled by a discontinuity in the equilibrium current, which is rounded off at finite temperatures. At finite temperature, the current is suppressed in the region of odd ground state parity between the two parity switches.\label{fig:cpr}}
\end{center}
\end{figure}

The current operator can be expanded in the eigenbasis of the linearized Hamiltonian by using Eq.~(\ref{eq:eigenmode_expansion}): 
\begin{align}\label{eq:current_eigenbasis}
j = &  \sum_{n} \left( \Gamma_{n}^{\dagger} \Gamma_{n}  -  \tfrac{1}{2}\right) \, j_{n,n}\\\nonumber
&  + \tfrac{1}{2} \sum_{n \neq m} (\Gamma_{n}^{\dagger} \Gamma_{m} \, j_{n,m}+ \Gamma_{n}^{\dagger} \Gamma_{m}^{\dagger} \, j_{n,\mathcal{P}m} + \textrm{H.c.})\,.
\end{align}
Here we have introduced the matrix elements of the current operator between BdG eigenstates,
\begin{align}
j_{n,m} &= e v\, \Phi_n^\dagger s_z \Phi_m\,,\\
j_{n,\mathcal{P}m} &= e v\, \Phi_n^\dagger s_z \mathcal{P}\Phi_m\,.
\end{align}
The diagonal matrix elements $j_{n,n}$ in Eq.~\eqref{eq:current_eigenbasis} give the dissipationless supercurrent.
Including, as already discussed, only the contribution from Andreev bound states to the sum in Eq.~\eqref{eq:current_eigenbasis}, we find
\begin{equation}\label{eq:equilibrium_current}
\av{j} = \left(n_1 -  \tfrac{1}{2}\right)\,j_{1,1}+\left(n_2-\tfrac{1}{2}\right)\,j_{2,2}\,,
\end{equation}
with $\av{\cdot}$ being the quantum expectation value and $n_n=\av{\Gamma^\dagger_n\Gamma_n}$ the occupation factors for the different quasiparticle states.
At thermal equilibrium with temperature $T$, $\av{\Gamma^\dagger_n\Gamma_n}_\textrm{eq}=f(E_n)$, where $f(E)=[1+\exp(E/k_BT)]^{-1}$ is the Fermi-Dirac distribution.
At $B=0$, the diagonal matrix elements have a simple analytic expression which can be computed using the wave function in Eq.~\eqref{eq:wavefzerofield} in Appendix~\ref{app:zero_field_bound_state_wavefunctions},
\begin{equation} \label{eq:current_diag_zerofield}
j_{1,1}=j_{2,2}= \frac{e}{2}\frac{\Delta_{0}^{2}\tau\sin\phi}{E_{A}(\phi)} \,, \quad (B=0)\,.
\end{equation} 
The result is independent of $\mu$ and $\alpha$, as long as the Andreev approximation is valid, see Sec.~\ref{sec:SolutionAtZeroField}.
Plugging the expression above into Eq.~\eqref{eq:equilibrium_current} immediately leads to the known expression for the Josephson current in a single channel weak link 
\begin{equation}\label{eq:av_j_B=0}
\av{j}_\textrm{eq} = \frac{e}{2\hbar}\frac{\Delta^2_0}{E_A(\phi)}\,\tau \sin\phi\,\tanh\left[\frac{E_A(\phi)}{2k_BT}\right]\,.
\end{equation}
In this zero field case, the fact that $j_{1,1} = j_{2,2}$ has the consequence that the Josephson current vanishes if the state of junction is one of the two odd parity states.
Namely, from Eq.~\eqref{eq:equilibrium_current} we see that $\av{j}=0$ if $j_{1,1} = j_{2,2}$ and $n_1+n_2=1$.
This is the so-called ``poisoned'' state of the junction \cite{zgirski2011,olivares2014}, in which one excess quasiparticle can completely block the passage of current.
Note that, if the junction has more than one pair of Andreev bound states, a single excess quasiparticle will not block the current completely, as there will be more contributions to the total equilibrium current.

The typical behavior of the equilibrium current-phase relation at finite magnetic field is illustrated in Fig.~\ref{fig:cpr}.
At small fields, the behavior is not qualitatively different from that of Eq.~\eqref{eq:av_j_B=0}.
At low temperatures the current-phase relation exhibit the skewed-sine shape typical of weak links, with the skewness being suppressed with increasing temperatures (see upper panel in Fig.~\ref{fig:cpr}).
The behavior is more interesting at higher fields, such that fermion parity switches occur as a function of the phase $\phi$, as in the Andreev spectrum in the upper right panel of Fig.~\ref{fig:abs_spectra}.
In this case, at $T=0$ the current exhibits a discontinuity in correspondence with each fermion parity switch (see bottom panel in Fig.~\ref{fig:cpr}).
At finite temperatures there is no discontinuity, but a remnant of the fermion parity switches remains in the behavior of the current phase relation close to $\phi=\pi$.
Finally, we mention that, as expected, the current model does not exhibit the anomalous Josephson effect (i.e., a finite supercurrent at $\phi=0$).
Indeed, for single-channel nanowire Josephson junction, it is known that the latter requires a component of the magnetic field to be aligned with the spin-orbit field \cite{nesterov2016}.

\section{Microwave absorption}
\label{sec:absorption}

In this section we study the microwave absorption spectrum of a short Josephson junction \cite{bergeret2010,kos2013,virtanen2013,bretheau2014,peng2016} (for the opposite case of a long junction, see also Refs.~\cite{dassonneville2013,ferrier2013,vayrynen2015}).
The microwave field is modeled as a monochromatic ac voltage drop $V(t) = V_0 \cos(\omega t)$ across the junction and is minimally coupled to the electronic field $\psi$.
This leads to the addition of the following time-dependent term to the Hamiltonian of Eq.~\eqref{eq:wire_model}:
\begin{equation}\label{eq:coupling_hamiltonian}
\delta H(t) = j\,(V_0/\omega)\,\sin(\omega t)\,.
\end{equation}
where $j$ is the current operator evaluated at the junction.
We assume that the perturbation $\delta H$ is small, $e V_0/\omega \ll 1$.
The form of the perturbation $\delta H(t)$ remains valid also after the spectrum linearization, s{}ince as we have discussed in the previous Section the current matrix elements at the position of the junction remain well defined and continuous, $j(0^-)=j(0^+)\equiv j$.
Using standard linear response theory, the expectation value of the current at time $t$ is determined by the response function $\chi(t) = -i\,\theta(t)\,\av{[j(t), j(0)]}_\textrm{eq}$,
\begin{equation}
\av{j(t)} = \av{j}_\textrm{eq}+\frac{V_0}{\omega}\int_{-\infty}^\infty\,\chi(t-t')\,\sin(\omega  t')\,\de t'\,.
\end{equation}
In the frequency domain, the response function $\chi(\omega)$ determines the admittance of the junction, $Y(\omega) = i\chi(\omega)/\omega$.
In turn, the real part of the admittance gives the absorption power $W$ of the microwave radiation, $W = \tfrac{1}{2}\,V_0^2\,\re Y(\omega)$ with $\omega>0$.
Using Eq.~(\ref{eq:current_eigenbasis}) to compute the response function, we find
\begin{widetext}
\begin{align}
\text{Re}\, Y(\omega) = & \frac{\pi}{\omega}\sum_{E_n \geq E_m}|j_{n,\mathcal{P}m}|^{2} \delta(\omega-(E_{m}+E_{n})) \left(1-f(E_{m})-f(E_{n})\right) \nonumber \\ 
 & +\frac{\pi}{\omega} \sum_{E_n \geq E_m}|j_{n,m}|^{2} \delta(\omega-(E_{n}-E_{m})) \left(f(E_{m})-f(E_{n})\right) +\dots \,. \label{eq:Yomega}
\end{align}
\end{widetext}
The first line in Eq.~(\ref{eq:Yomega}) corresponds to transitions where two quasiparticles are created by breaking a Cooper pair and occupy two energy levels with energies $E_n$ and $E_m$.
The second line corresponds to transitions where a single quasiparticle with energy $E_n$ is excited into a higher state with energy $E_m$.
We will refer to these two types of transition respectively as the ``even'' or ``odd'' ones, since they are distinguished by the parity of the number of quasiparticles involved. Note that only transitions in which initial and final states have the same fermion parity are allowed. Transitions between the discrete states, which are accounted for in Eq.~(\ref{eq:Yomega}), produce sharp maxima in the frequency dependence of the absorption coefficient. 
The omitted terms in the admittance, indicated by dots in Eq.~(\ref{eq:Yomega}), involve unbound quasiparticle states and result in an absorption continuum. 

We shall consider low frequencies $\omega < 2\Delta$, focusing on the transitions between the Andreev bound states.
Indeed, at these low frequencies the excitation of Andreev states are the only possible resonant processes (unless the system is close to the critical point separating topological and trivial phases, a case treated in Ref.~\cite{tewari2012}).
Transitions between possible above-gap non-equilibrium quasiparticles are very weak and do not result in a sharp absorption line, so we will ignore them.
In the case under consideration of a single-channel short junction, with only two Andreev states with energies $E_1$ and $E_2$, there is only one relevant term in each sum in Eq.~(\ref{eq:Yomega}).
These terms correspond to the two allowed transitions depicted in the bottom right panel of Fig.~\ref{fig:diagram_spectrum}: the pair excitation $\ket{V}\to \ket{P}$ with frequency $\omega_\textrm{even}=E_1 + E_2$ and the single particle excitation $\ket{1}\to\ket{2}$ with frequency $\omega_\textrm{odd}=E_2-E_1$. 
As mentioned in the introduction, we call these ``even'' and  ``odd'' transitions, respectively. 
The matrix elements $j_{2,\mathcal{P}1}(0^{-})$ and $j_{2,1}(0^{-})$ determine the strengths of these transitions; their dependence on the system parameters is discussed next in detail, first for the even transition and then for the odd one.

\begin{figure}[t!]
\begin{center}
\includegraphics[width=\columnwidth]{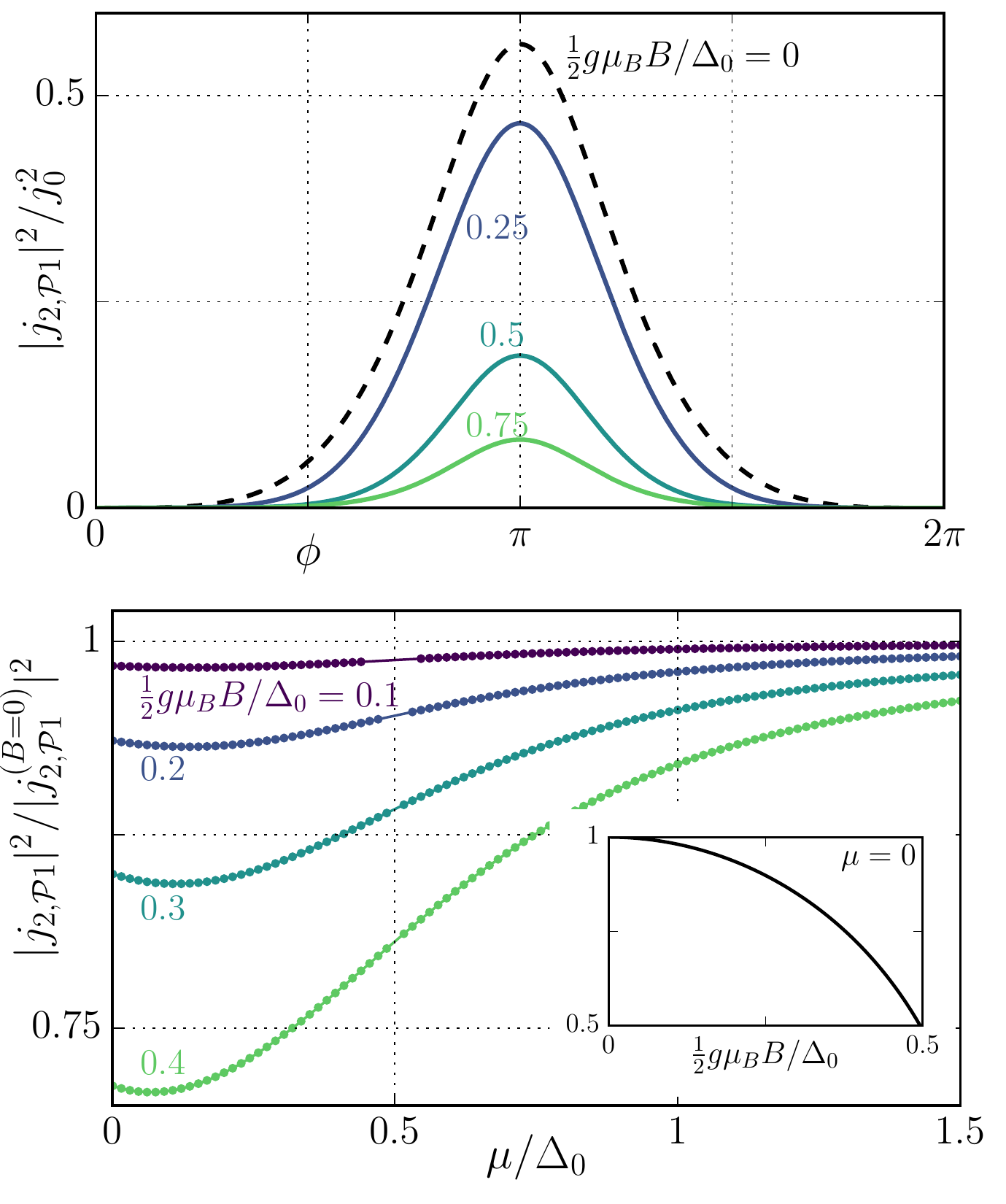}
\caption{\emph{Top panel:} Dependence of the square of the current matrix element $\abs{j_{2,\mathcal{P}1}}^2$, which determines the visibility of the even transition $\ket{V}\to \ket{P}$, on the phase $\phi$ for different magnetic fields $B$, in the regime $\mu\gg \Delta_0, m\alpha^2$. Other parameters are the same as Fig.~\ref{fig:cpr}: $\alpha k_F/\Delta_0=0.5, \tau=0.75$. As in Fig.~\ref{fig:cpr}, the current is in units of $j_0=e\Delta_0/\hbar$. The $B=0$ curve (black dashed line) is given by Eq.~\eqref{eq:j2P1ZeroField}, the rest of the curves are determined numerically. \emph{Bottom panel:} Dependence of the same current matrix element $\abs{j_{2,\mathcal{P}1}}^2$ on chemical potential $\mu$ (main figure) and magnetic field $B$ (inset) for the case $m\alpha^2\gg \Delta_0, \mu$ and for $\phi=\pi$, $\tau=0.75$. The chemical potential dependence is given for different values of the magnetic field, see labels close to each curve. The current matrix elements are normalized by their zero field value, Eq.~\eqref{eq:j2P1ZeroField}, which is independent of $\mu$.\label{fig:j2P1}}
\end{center}
\end{figure}

\subsection{Visibility of the even transition $\ket{V}\to\ket{P}$}
\label{sec:absorption_even}

We start by discussing the case $B=0$.
Again by using the wave functions in Eq.~(\ref{eq:wavefzerofield}) of Appendix~\ref{app:zero_field_bound_state_wavefunctions}, we find the analytical expression for the relevant current matrix element
\begin{equation} \label{eq:j2P1ZeroField}
\abs{j_{2,\mathcal{P}1}}^2= e^2 (1-\tau)\,\tau^2\,\sin^4(\phi/2)\,(\Delta_0^4/E_A^2)
\end{equation}
Equation~(\ref{eq:j2P1ZeroField}) was previously derived in Ref.~\cite{kos2013}using a tunneling Hamiltonian formalism, which is in agreement with our current method based on the transfer matrix.
Just like the $B=0$ Andreev energy $E_A$, it is independent on the chemical potential $\mu$ and the spin-orbit coupling $\alpha$ and it generalizes to the case of multiple transport channels with different transparencies.
Note that $\abs{j_{2,\mathcal{P}1}}^2$ vanishes for $\tau=1$: the absence of scattering at the junction prevents the excitation of the Andreev bound states since in this case the current is a diagonal operator in the eigenbasis of Eq.~\eqref{eq:current_eigenbasis}.
In the presence of scattering, the Andreev bound states are superpositions of different current eigenstates and microwave-induced transitions become possible \cite{zazunov2005}.
Equation~\eqref{eq:j2P1ZeroField} has a maximum at $\phi=\pi$, corresponding to the point of greater visibility of the absorption spectral line.
The visibility vanishes for small phases.
This behavior is in agreement with experiment both in case of nanowire Josephson junctions \cite{vanwoerkom2016} as well as other types of weak links \cite{bretheau2013,bretheau2014}.

The dependence of $j_{2,\mathcal{P}1}$ on magnetic field can be determined by finding the wave functions of the Andreev bound states numerically via Eq.~\eqref{eq:wavefunction}.
We find that a finite magnetic field suppresses the magnitude of the current matrix element, while maintaining its phase dependence qualitatively similar to that described by Eq.~\eqref{eq:j2P1ZeroField}, see upper panel of Fig.~\eqref{fig:j2P1}.
The decrease of $j_{2,\mathcal{P}1}$ with increasing magnetic field is slow in both regimes $m\alpha^2 \gg \mu_0, \Delta_0$ (see the inset of the bottom panel in Fig.~\ref{fig:j2P1}) and $\mu\gg \Delta_0, m\alpha^2$ (see Fig.~\ref{fig:ReY_vs_B_high_mu}).
We attribute this decrease to the suppression of the proximity-induced gap $\Delta(B)$ with $B$ (see Fig.~\ref{fig:continuum_gap}), which makes the Andreev bound states less tightly confined to the junction and thereby decreases the effective coupling to microwaves, $j_{2,\mathcal{P}1} \sim \Delta(B)$. 
Finally, at finite fields, the current matrix elements also acquires a weak dependence on the chemical potential, as illustrated in the bottom panel of Fig.~\eqref{fig:j2P1}.

\begin{figure}[t!]
\begin{center}
\includegraphics[width=\columnwidth]{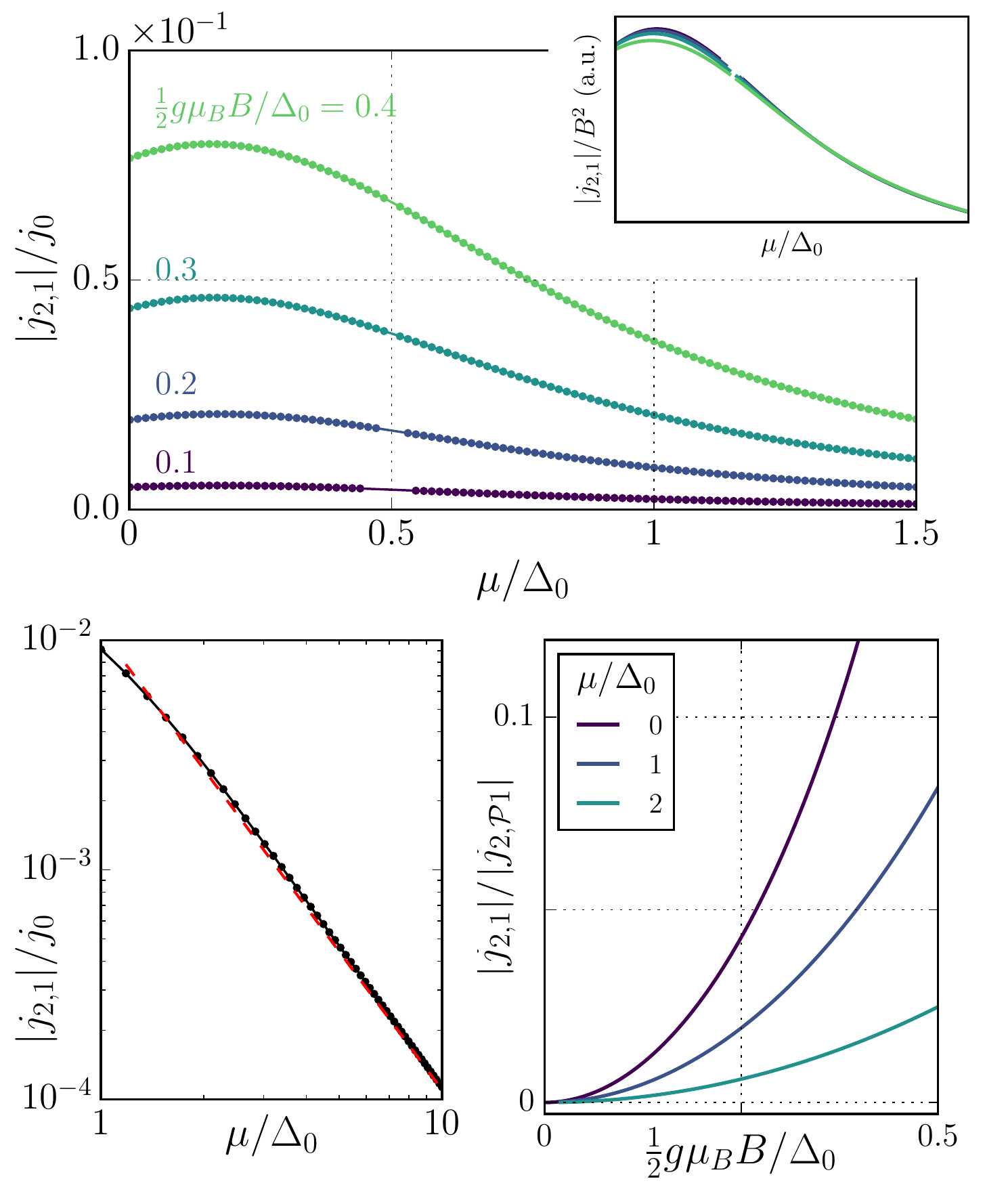}
\caption{\emph{Top panel:} dependence of the magnitude of the current matrix element $\abs{j_{2,1}}$, which determines the visibility of the odd transition $\ket{1}\to \ket{2}$, on the chemical potential $\mu$ for different magnetic fields $B$, in the regime $m\alpha^2\gg \Delta_0, \mu$. Other parameters: $\phi=\pi, \tau=0.75$. As in previous figures, $j_0=e\Delta_0/\hbar$. In the inset, the data for each curve is rescaled by $B^2$, to show the reasonable agreement with the $\abs{j_{2,1}}\propto B^2$ behavior, particularly when $\mu\gg \Delta_0$. The few missing numerical data points at $\mu/\Delta_0\approx 0.5$ are in correspondence with the narrow chemical potential interval in which the Andreev level g-factor in Eq.~\eqref{eq:g_counterpropagating} vanishes in the limit $\mu/m\alpha^2,\Delta_0/m\alpha^2\to0$. In this case, the energy levels cannot be resolved numerically. \emph{Bottom panels:} On the left, we show the dependence of continuation of $\abs{j_{2,1}}$ on $\mu$ at large values $\mu/\Delta_0$ for $\tfrac{1}{2}g\mu_B B/\Delta_0=0.2$, in log-log scale. The dashed red line has slope -2, demonstrating $\abs{j_{2,1}}\propto (\Delta_0/\mu)^2$ for $\mu/\Delta_0\gtrsim 1$. On the right, we show the magnetic field dependence of the ratio $\abs{j_{2,1}}/\abs{j_{2,\mathcal{P}1}}$ for different values of $\mu$, illustrating that the odd transition is much less visible than the even transition.
\label{fig:j21}}
\end{center}
\end{figure}

\subsection{Visibility of the odd transition $\ket{1}\to\ket{2}$}
\label{sec:absorption_odd}

Without magnetic field, $B=0$, the current matrix element associated with the odd transition (which has anyway zero frequency) vanishes: $j_{2,1}=0$.
This is due to the fact that the zero-field Andreev bound states have opposite spin [see Eqs.~\eqref{eq:wavefzerofieldup} and \eqref{eq:wavefzerofielddown}], while the perturbation Hamiltonian \eqref{eq:coupling_hamiltonian} preserves spin.
As the magnetic field is increased from zero, the odd transition may become visible depending on the spin-orbit coupling strength.
If spin-orbit is absent (or negligible), the two Andreev bound states would develop an opposite spin polarization in the presence of a Zeeman field: therefore, again due to the spin selection rule, the odd transition would remain forbidden.
In the presence of both spin-orbit coupling and magnetic field, however, this spin selection rule is no longer applicable: the two Andreev bound states would have a non-zero spin overlap and one may in general expect a non-vanishing matrix element.
Indeed, we determine numerically that $j_{2,1}\neq 0$ at finite $B$ in the presence of spin-orbit coupling.
Importantly, even in this case we find that $\abs{j_{2,1}}/\abs{j_{2,\mathcal{P}1}}\lesssim 0.1$, see the bottom right panel of Fig.~\ref{fig:j21}.
Hence, despite not being forbidden, the dim odd transition may be still much more difficult to observe with respect to the bright even transition.
We now discuss the dependence of $j_{2,1}$ on the system parameters.

\begin{figure}[t!]
\begin{center}
\includegraphics[width=\columnwidth]{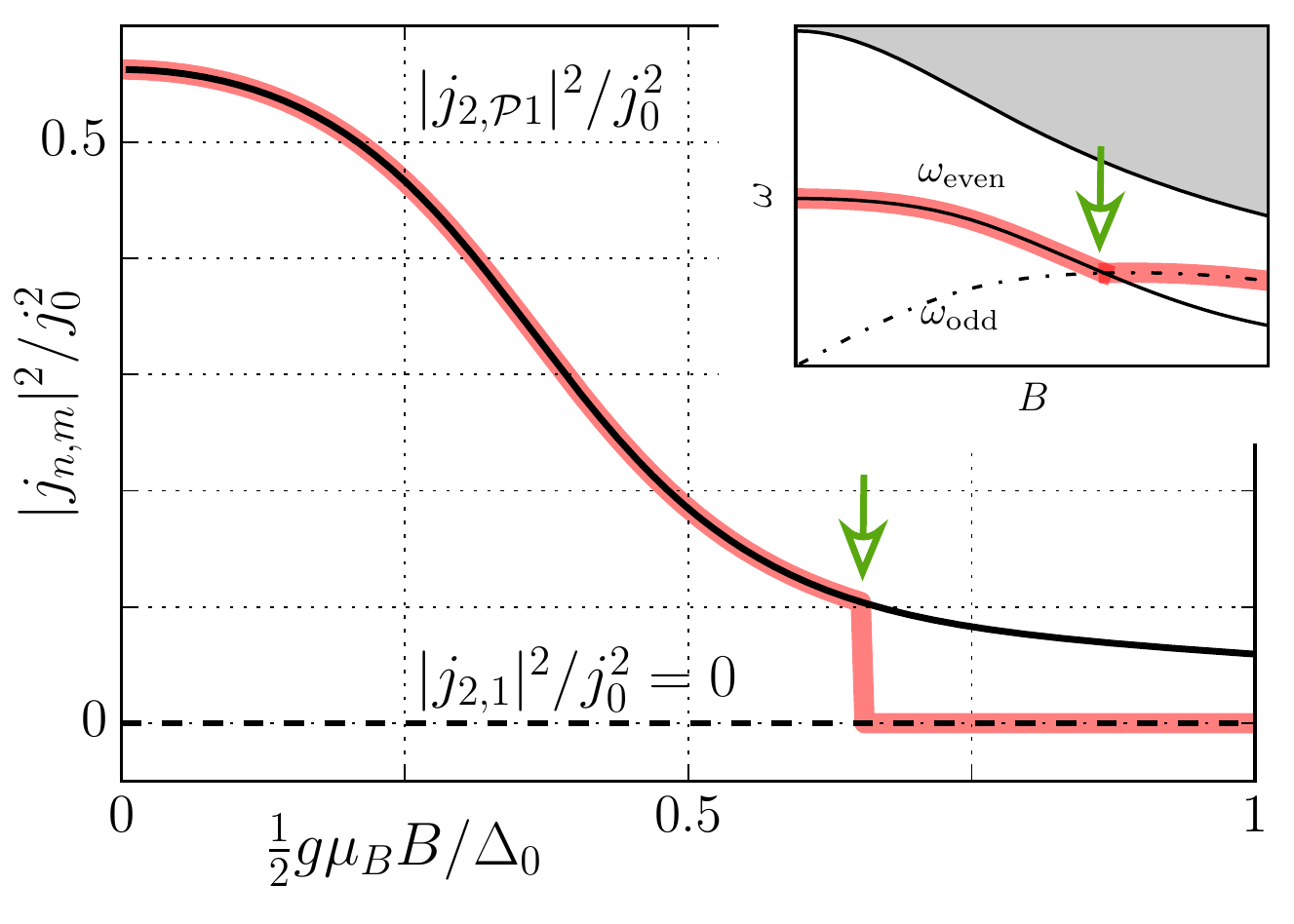}
\caption{Current matrix elements versus magnetic field in the regime $\mu\gg m\alpha^2, g\mu_B B, \Delta_0$, computed for the same parameter values as the left panel of Fig.~\ref{fig:abs_spectra}. The matrix elements $\abs{j_{2,\mathcal{P}1}}^2$ (solid line) and $\abs{j_{2,1}}^2$ (dashed line) determine the visibility of the even and odd transition lines in the absorption spectrum of the junction. The magnetic field dependence of the corresponding frequencies $\omega_\textrm{even}$ and $\omega_\textrm{odd}$, already shown in Fig.~\ref{fig:abs_spectra}, is reproduced in the inset. A fermion parity occurs at the value of the magnetic field $B=B_\textrm{sw}$ marked by the green arrow, where $\omega_\textrm{even}=\omega_\textrm{odd}$. The thick red shadow follows the frequency (inset) and visibility (main panel) of the absorption line which would be measured at low temperatures, i.e. assuming that the microwave absorption always excites the junction from its ground state. The visibility of the absorption line, in this case, drops drastically to zero in correspondence with the kink in the transition frequency at $B=B_\textrm{sw}$.}
\label{fig:ReY_vs_B_high_mu}
\end{center}
\end{figure}

As in the case of the even transition, the current matrix element $\abs{j_{2,1}}$ has a maximum when $\phi=\pi$ and vanishes for small phase differences; in what follows, we focus on the peak value.
The dependence of $\abs{j_{2,1}}$ on chemical potential is shown in the top panel of Fig.~\ref{fig:j21} for different values of $B$.
The current matrix elements is non-zero at $\mu=0$, it grows slowly and it reaches a maximum at a small value of $\mu/\Delta_0$ before decreasing again.
After this point, we find that $\abs{j_{2,1}}\propto (\Delta_0/\mu)^2$ when $\mu/\Delta_0\gg 1$, as shown in the bottom left panel of Fig.~\ref{fig:j21}.
These considerations are valid when $m\alpha^2\gg \Delta_0, \mu$.
The suppression of $\abs{j_{2,1}}$ for $\mu\gg \Delta_0$ in this regime matches the numerical results that we obtain for $\mu\gg m\alpha^2$, where we find that the matrix element is zero (within numerical precision) at any value of the spin-orbit coupling.
Finally, the numerical data indicate that the current matrix elements grows quadratically in $B$ at small fields: $\abs{j_{2,1}}\propto B^2$, see the inset in the top panel in Fig.~\ref{fig:j21}.

The smallness of the current matrix element $j_{2,1}$ has important consequences for the Andreev spectroscopy of the junction at low temperatures, in case the junction undergoes a fermion parity switch.
For instance, suppose that the magnetic field is sweeped from a value $B<B_\textrm{sw}$ to a value $B>B_\textrm{sw}$, as in the left panel of Fig.~\ref{fig:abs_spectra}.
This change of magnetic field will be accompanied by a dramatic decrease in the visibility of the absorption line corresponding to the even transition at frequency $\omega_\textrm{even}$.
Indeed, for $B>B_\textrm{sw}$ the ground state of the junction is the odd parity state $\ket{1}$, and at low temperatures $k_B T\ll E_1$ the occupation probability of the even parity state $\ket{V}$ is negligible.
The low occupation probability of the state $\ket{V}$ and the smallness of the matrix element $j_\textrm{2,1}$ combine to yield a dramatic dimming of the absorption line taking place at $B=B_\textrm{sw}$ (see Fig.~\ref{fig:ReY_vs_B_high_mu}).

\section{Conclusions}

In this work, we have investigated several consequences of the competition between Zeeman and spin-orbit couplings on the Andreev bound states in semiconducting nanowire Josephson junctions.
Overall, as one may have expected, spin-orbit coupling tends to reduce the effect of the Zeeman coupling on the Andreev bound states.
We have seen several examples of this general trend.
First, as discussed in Sec.~\ref{sec:g-factor}, spin-orbit coupling tends to suppress the g-factor $g_A$ of the Andreev bound states, potentially resulting in very small energy splittings of the Andreev doublet for small magnetic fields.
The measurement of $g_A$, in tunneling or supercurrent spectroscopy experiments, may allow one to estimate the strength of the spin-orbit coupling.
Second, spin-orbit coupling also suppresses the occurrence of fermion parity switches in the topologically trivial phase of the nanowire (see Sec.~\ref{sec:fermi_level_crossings} and Fig.~\ref{fig:b_sw_vs_spin-orbit}).
As discussed at the end of Sec.~\ref{sec:absorption}, fermion parity switches should be easily detectable since they are accompanied by a drastic dimming of the absorption spectrum.
The knowledge of the switching field $B_\textrm{sw}$ at which fermion parity switches take place can also be used to infer the strength of the spin-orbit coupling.
Finally, spin-orbit coupling prevents the occurrence of level crossings between the Andreev bound states and the continuum.
Combined with the suppression of the proximity-induced energy gap in a magnetic field, this leads to a non-monotonic dependence of the Andreev bound state energies on $B$, see the left panel of Fig.~\ref{fig:abs_spectra}.
The bending of the Andreev level $E_2$ due to the repulsion from the continuum causes a slow decrease of the even transition frequency in magnetic field, $\omega_\textrm{even}(0)-\omega_\textrm{even}(B)\propto B^2$ for small $B$.

Our theoretical results are in good agreement with several aspects of the existing experimental data which motivated the development of the work presented here \cite{vanwoerkom2016}.
In particular, we elucidated that the quadratic suppression of $\omega_\textrm{even}$ with increasing magnetic field can be understood in terms of the interplay of Zeeman and spin-orbit coupling.
The occurrence of fermion parity switches is also compatible with the observation that the even transition visibility vanished at a field larger than $300$ mT.
This threshold can be well understood within our theory assuming reasonable values of $g$ and $\alpha$ \cite{vanwoerkom2016}.
At the experimental level, it would be very valuable to study directly the single-particle energy spectrum via either tunneling or supercurrent spectroscopy.
This would allow a measurement of the Andreev bound state g-factor $g_A$ as well as a precise determination of the switching field $B_\textrm{sw}$, both of which can be directly compared to our theory.

Our results are all based on the one-dimensional nanowire model of Eq.~\eqref{eq:wire_model} treated within the Andreev approximation, i.e. by linearizing the normal state dispersion.
This approximation amounts to neglecting the normal reflection amplitude in favor of the Andreev reflection amplitude when considering the two interfaces of the S-N-S junction.
It requires that either the chemical potential $\mu$ or the spin-orbit energy $m\alpha^2$ are much larger than the induced superconducting gap $\Delta_0$.
As an extension of this work, it may be valuable to relax the Andreev approximation.
In particular, it would be interesting to study the Andreev spectrum of the model in the regime $\mu\ll\ m\alpha^2, \Delta_0$ and $m\alpha^2\sim \Delta_0$, which may be relevant for the Majorana applications of the semiconducting nanowires.
This may give more accurate predictions for the Andreev g-factor $g_A$ and the current matrix elements in the regime of low chemical potential.

It will also be important to extend this work beyond the model of Eq.~\eqref{eq:wire_model}, in order to capture more accurately the complexity of real devices.
Nanowire junctions may naturally host more than one transport channel, and physical effects not included in this work, such as the orbital effect of the magnetic field, may have an important influence on the Andreev bound state properties.
In particular, the orbital effect of the magnetic field provides an additional contribution to the reduction of $\omega_\textrm{even}$.
Although this contribution could be heuristically ruled out to be the dominant one in the current nanowire experiments, it would be important to have quantitative theoretical estimates.

Finally, the magnetic field dependence of the absorption spectrum in the presence of multiple transport channels stands out as a particularly interesting avenue for future research, both theoretically and experimentally.
In such a situation, a new type of low-frequency transitions may become visible, in which a Cooper pair is excited to a pair of Andreev levels belonging to different transport channels.
In the topological phase, we expect that these inter-channel transitions can carry a signature of the Majorana bound states in the form of a kink in the phase dependence of the absorption spectrum, similar to the effect predicated in long Josephson junctions \cite{vayrynen2015}.
Notably, this type of measurement is not limited by stringent requirements on fermion parity relaxation times, as opposed to other signatures of topological Josephson junctions.

\acknowledgments

We acknowledge stimulating discussions with A.~Geresdi, D.J.~van~Woerkom, H.~Pothier, R.~Lutchyn, T.~Hyart and S.~Park. BvH was supported by ONR Grant Q00704, JV and LG acknowledge the support by NSF DMR Grant No. 1603243.

\bibliography{references}

\appendix

\section{Bound state wave functions at zero field}
\label{app:zero_field_bound_state_wavefunctions}

Following the procedure outlined in the main text, we find the two following bound state wave functions $\Phi_1(x)$ and $\Phi_2(x)$ at $B=0$. They are a tensor product of a position-dependent part and a position-independent spinor in spin grading:
\begin{align}
\Phi_1(x) &= \e^{i\phi\sgn(x) \tau_z/4}\,\Phi_A(x)\,\otimes\,\chi_\up\,  \label{eq:wavefzerofieldup} \\
\Phi_2(x) &= \e^{i\phi\sgn(x) \tau_z/4}\,\Phi_A(x)\,\otimes\,\chi_\down\,.  \label{eq:wavefzerofielddown}
\end{align}
Here $\chi_{\up}=(1,0)^T$ and $\chi_\down=(0,1)^T$ are the eigenspinors of $\sigma_z$, and $\Phi_A(x)$ is a space-dependent vector in Nambu and left/right gradings:
\begin{widetext}
\begin{equation} \label{eq:wavefzerofield}
\Phi_A(x) = \frac{1}{2\,\xi_A^{1/2}}\frac{1}{[E_A\,(E_A-\Delta_0\sqrt{\tau}\cos\phi/2)]^{1/2}}\,\e^{-i\alpha k_F x/v_F}\,\e^{-\abs{x}/\xi_A}\,\left[
\begin{array}{c}
- \sgn (x) i \e^{i\gamma\,\theta(-x)}\e^{i\beta\,\theta(x)}\,(E_A-\Delta_0\sqrt{\tau}\cos\phi/2)\\
\e^{i\beta\,\theta(-x)}\e^{i\gamma\,\theta(x)}\,\Delta_0\,\sqrt{1-\tau}\\
- \sgn (x) i \e^{i\gamma\,\theta(-x)}\e^{i\beta\,\theta(-x)}\,(E_A-\Delta_0\sqrt{\tau}\cos\phi/2)\\
\e^{i\beta\,\theta(x)}\e^{i\gamma\,\theta(x)}\,\Delta_0\,\sqrt{1-\tau}
\end{array}
\right]
\end{equation}
with:
\begin{equation}
E_A =  \Delta_0 [1-\tau\sin^2(\phi/2)]^{1/2}\;,\;\beta = \arccos(E_A/\Delta)\;,\;\gamma = \arccos(\sqrt{\tau})\;,\;\xi^{-1}_A = \frac{1}{v_F}\,\sqrt{\Delta_0^2-E_A^2}\,.
\end{equation}
The expression above is valid for $\mu\gg \Delta_0, m\alpha^2$ and in the phase interval $\phi\,\in\,[0,2\pi]$. The wave functions for negative phase can be determined by applying the time-reversal symmetry operator $is_x \sigma_y \mathcal{K}$. In the opposite regime $m\alpha^2\gg \mu, \Delta_0$, the wave functions are identical except that the oscillating term $\e^{-i\alpha k_F/v_F}$ is replaced by $\e^{-i\mu x/\alpha s_z}$ and that $\alpha$ replaces $v_F$ in the expression for the coherence length $\xi_A$ of the bound state. The wave functions above are properly normalized to unity: to see this, it is convenient to use the relation $\Delta_0^2\,(1-\tau)=(E_A-\Delta_0\sqrt{\tau}\cos\phi/2)\,(E_A+\Delta_0\sqrt{\tau}\cos\phi/2)$.

\section{Derivation of the different contributions to the g-factor}
\label{app:derivation_gfactor}

In the main text, Secs.~\ref{subsec:linearization_mu} and \ref{subsec:linearization_mu=0}, we linearized the spectrum in two
limits of either large $\mu$ or large $m\alpha^{2}$. In this Appendix,
we show that both limits can be obtained from a single linearization
which is valid on a strip of width $\mu+m\alpha^{2}$ around the Fermi
level. This linearization is achieved by a projection
\begin{flalign}
\psi(x) & =e^{-imx\,(\sqrt{\alpha^{2}+v_{F}^{2}}+\alpha\sigma_{z})}\,\psi_{L}(x)+e^{imx\,(\sqrt{\alpha^{2}+v_{F}^{2}}-\alpha\sigma_{z})}\,\psi_{R}(x)\label{eq:linearization_general}
\end{flalign}
where the fields $\psi_{L,R}$ are slowly varying. For example, the kinetic term in the Hamiltonian density becomes 
\begin{equation}
\psi(x)^{\dagger}\left(-\frac{\partial_{x}^{2}}{2m}-i\alpha\partial_{x}\,\sigma_{z}-\mu\right)\tau_{z}\psi(x)=-i\sqrt{\alpha^{2}+v_{F}^{2}}\Psi^{\dagger}(x)s_{z}\tau_{z}\partial_{x}\Psi(x) + \textrm{oscillating terms}\,.
\end{equation}
We used here $\mu=\frac{1}{2}mv_{F}^{2}$. 
When we project the Zeeman term $-\frac{1}{2}g\mu_{B}B\psi^{\dagger}\sigma_{x}\psi$
to low energies using Eq.~(\ref{eq:linearization_general}), we obtain
two terms, 
\begin{equation}
-\frac{1}{2}g\mu_{B}B\psi^{\dagger}\sigma_{x}\psi=\Psi^{\dagger}\left(O_{\rightrightarrows}+O_{\rightleftarrows}\right)\Psi\,,
\end{equation}
where 
\begin{equation}
O_{\rightrightarrows}=-\frac{1}{2}g\mu_{B}Be^{2im\alpha x\sigma_{z}}\sigma_{x}\,,\quad O_{\rightleftarrows}=-\frac{1}{2}g\mu_{B}Be^{-2imx(s_{z}\sqrt{\alpha^{2}+v_{F}^{2}}-\sigma_{z}\alpha)}s_{x}\sigma_{x}\,.
\end{equation}
The first term couples co-propagating states only and it is important when spin-orbit strength is not too large, $m\alpha^{2}\lesssim\Delta_{0}$.
It leads to Eq.~(\ref{eq:g_copropagating}) of the main text, which can be derived by evaluating the matrix elements of $O_{\rightrightarrows}$ by using the wave functions from Eq.~\eqref{eq:wavefzerofield}.

The second term, $O_{\rightleftarrows}$, mixes counter-propagating states and therefore it only contributes to the g-factor in the presence of scattering at the junction.
Furthermore, it is important only for states near $k=0$ and when $\mu\ll m\alpha^{2}$, in which case the oscillating exponent vanishes.
States belonging to the outer branches at finite momentum have opposite $s_{z}$ and $\sigma_{z}$ eigenvalues, and in this case the $O_{\rightleftarrows}$ term oscillates fast and is negligible.
In the limit $\mu/m\alpha^{2}\to 0$, we thus obtain
\begin{equation}
O_{\rightleftarrows}=-\frac{1}{4}g\mu_{B}B(s_{x}\sigma_{x}-s_{y}\sigma_{y})\,.
\end{equation}
After calculating the matrix elements of $O_{\rightleftarrows}$ with the wave functions from Appendix~\ref{app:zero_field_bound_state_wavefunctions},
we find Eq.~(\ref{eq:g_counterpropagating}) of the main text.
Note that correction of order $\mu/m\alpha^2$ to the matrix elements of $O_{\rightleftarrows}$ cannot be reliably computed within the linearized Hamiltonian.
\end{widetext}

\end{document}